\documentstyle[12pt]{article}
\begin{document}

%******************************************************************
%\include{m00}
%%%%% MACROS FOR EEM3 DOCUMENT %%%%%
\def\tdot{\relax
\mbox{\raise2.6pt%
\hbox{\hspace*{2pt}\.{}\hspace{-.8pt}\.{}\hspace{-.8pt}\.{}\hspace{-8.0pt}}}}

\def\delt{\partial_{xx}+\partial_{yy}+\partial_{zz}}
\def\delw{\partial_{xx}+\partial_{yy}}

\def\A{{\cal A}}
\def\Bo{{\cal B}^1}
\def\Bw{{\cal B}^2}
\def\Bt{{\cal B}^3}
\def\C{{\cal C}}
\def\D{{\cal D}}
\def\E{{\cal E}}
\def\F{{\cal F}}
\def\G{{\cal G}}
\def\uo{{\cal I}}
\def\J{{\cal J}}
\def\jp{{\cal J}_+}
\def\jm{{\cal J}_-}
\def\K{{\cal K}}
\def\kp{{\cal K}_+}
\def\km{{\cal K}_-}
\def\L{{\cal L}}
\def\M{{\cal M}}
\def\O{{\cal O}}
\def\S{{\cal S}}
\def\U{{\cal U}}
\def\W{{\cal W}}
\def\R{{\cal R}}
\def\X{{\cal X}}
\def\Y{{\cal Y}}

\def\tsL{{\tilde{\L}}}
\def\tsJ{{\tilde{\J}}}
\def\tsK{{\tilde{\K}}}
\def\tL{{\tilde{L}}}
\def\tJ{{\tilde{J}}}
\def\tK{{\tilde{K}}}

\def\bZp{{\bf Z}^+}
\def\bR{{\bf R}}
\def\bC{{\bf C}}
\def\bN{{\bf N}}
\def\bZ{{\bf Z}}

\def\swp{S_{2\scriptstyle{+}}}
\def\swm{S_{2\scriptstyle{-}}}
\def\swpm{S_{2\scriptstyle{\pm}}}
\def\sswp{\S_{2\scriptstyle{+}}}
\def\sswm{\S_{2\scriptstyle{-}}}
\def\sswpm{\S_{2\scriptstyle{\pm}}}

\def\spm{\scriptstyle{\pm}}

\def\stp{S_{3\scriptstyle{+}}}
\def\stm{S_{3\scriptstyle{-}}}
\def\stpm{S_{3\scriptstyle{\pm}}}

\def\dzo{\partial_{\zeta_1}}
\def\dzw{\partial_{\zeta_2}}
\def\dzp{\dzo+i\dzw}
\def\dzm{\dzo-i\dzw}
\def\zpz{\zeta_1+i\zeta_2}
\def\zmz{\zeta_1-i\zeta_2}

\def\adg{a^{\dagger}}
\def\bdg{b^{\dagger}}
\def\cdg{c^{\dagger}}
\def\Qdg{Q^{\dagger}}

\def\pt{{\varphi}}
\def\bx{\bar{\xi}}
\def\dln{x\partial_x+y\partial_y}
\def\dv{\dot{\varphi}}
\def\ddv{\ddot{\varphi}}

\newcommand{\ketf}[4]{\mbox{$\vert#1,#2,#3,#4\rangle$}}
\newcommand{\ketfsc}[4]{\mbox{$\vert#1,#2;#3,#4\rangle$}}
\newcommand{\kett}[3]{\mbox{$\vert#1,#2,#3\rangle$}}
\newcommand{\kettsc}[3]{\mbox{$\vert#1,#2;#3\rangle$}}
\newcommand{\ketw}[2]{\mbox{$\vert#1,#2\rangle$}}
\newcommand{\bra}[3]{\mbox{$\langle#1,#2,#3\vert$}}

\newcommand{\lfrac}[2]{\mbox{${#1}\over{#2}$}}

\def\rep{{representation}}
\def\reps{{representations}}
\def\irrep{{irreducible representation}}
\def\irreps{{irreducible representations}}
\def\unirrep{{unitary irreducible representation}}
\def\unirreps{{unitary irreducible representations}}

%**************************************************************************
%\include{m0}
\font\tenrm=cmr10

\def\al{\alpha}
\def\be{\beta}
\def\ga{\gamma}
\def\de{\delta}
\def\ep{\epsilon}
\def\ve{\varepsilon}
\def\ze{\zeta}
\def\et{\eta}
\def\th{\theta}
\def\vt{\vartheta}
\def\io{\iota}
\def\ka{\kappa}
\def\la{\lambda}
\def\vpi{\varpi}
\def\rh{\rho}
\def\vr{\varrho}
\def\si{\sigma}
\def\vs{\varsigma}
\def\ta{\tau}
\def\up{\upsilon}
\def\ph{\phi}
\def\vp{\varphi}
\def\ch{\chi}
\def\ps{\psi}
\def\om{\omega}
\def\Ga{\Gamma}
\def\De{\Delta}
\def\Th{\Theta}
\def\La{\Lambda}
\def\Si{\Sigma}
\def\Up{\Upsilon}
\def\Ph{\Phi}
\def\Ps{\Psi}
\def\Om{\Omega}
\def\mn{{\mu\nu}}
\def\cl{{\cal L}}
\def\fr#1#2{{{#1} \over {#2}}}
\def\prt{\partial}
\def\ap{\al^\prime}
\def\apt{\al^{\prime 2}}
\def\apth{\al^{\prime 3}}
\def\vev#1{\langle {#1}\rangle}
\def\vev{\langle {\phi}\rangle}
\def\bra#1{\langle{#1}|}
\def\ket#1{|{#1}\rangle}
\def\bracket#1#2{\langle{#1}|{#2}\rangle}
\def\expect#1{\langle{#1}\rangle}
\def\sbra#1#2{\,{}_{{}_{#1}}\langle{#2}|}
\def\sket#1#2{|{#1}\rangle_{{}_{#2}}\,}
\def\sbracket#1#2#3#4{\,{}_{{}_{#1}}\langle{#2}|{#3}\rangle_{{}_{#4}}\,}
\def\sexpect#1#2#3{\,{}_{{}_{#1}}\langle{#2}\rangle_{{}_{#3}}\,}
\def\half{{\textstyle{1\over 2}}}
\def\frac#1#2{{\textstyle{{#1}\over {#2}}}}
\def\ni{\noindent}
\def\lsim{\mathrel{\rlap{\lower4pt\hbox{\hskip1pt$\sim$}}
    \raise1pt\hbox{$<$}}}
\def\gsim{\mathrel{\rlap{\lower4pt\hbox{\hskip1pt$\sim$}}
    \raise1pt\hbox{$>$}}}
\def\sqr#1#2{{\vcenter{\vbox{\hrule height.#2pt
         \hbox{\vrule width.#2pt height#1pt \kern#1pt
         \vrule width.#2pt}
         \hrule height.#2pt}}}}
\def\square{\mathchoice\sqr66\sqr66\sqr{2.1}3\sqr{1.5}3}

\newcommand{\beq}{\begin{equation}}
\newcommand{\eeq}{\end{equation}}
\newcommand{\bea}{\begin{eqnarray}}
\newcommand{\eea}{\end{eqnarray}}
\newcommand{\rf}[1]{(\ref{#1})}

\titlepage

\begin{flushright}
{IUHET 243\\}
{LA-UR-93-206\\}
%{hep-ph/9303068\\}
{January 1993\\}
\end{flushright}
\vglue 1cm

\begin{center}
{{\bf SUPERSYMMETRY AND A TIME-DEPENDENT LANDAU SYSTEM \\}
\vglue 1.0cm
{V. Alan Kosteleck\'y\\}
\medskip
{\it Physics Department\\}
\smallskip
{\it Indiana University\\}
\smallskip
{\it Bloomington, IN 47405, U.S.A.\\}
\vglue 0.5cm
{V. I. Man'ko\\}
\medskip
{\it P. N. Lebedev Physical Institute\\}
\smallskip
{\it Russian Academy of Science\\}
\smallskip
{\it Moscow 117924, Russian Federation.\\}
\vglue 0.5cm
{Michael Martin Nieto\\}
\medskip
{\it Theoretical Division\\}
\smallskip
{\it Los Alamos National Laboratory\\}
\smallskip
{\it University of California\\}
\smallskip
{\it Los Alamos, NM 87545, U.S.A.\\}
\vglue 0.5cm
{D. Rodney Truax\\}
\medskip
{\it Department of Chemistry\\}
\smallskip
{\it University of Calgary\\}
\smallskip
{\it Calgary, Alberta, Canada T2N 1N4.\\}

}

\end{center}

\vglue 0.3cm

{\rightskip=3pc\leftskip=3pc\noindent%\tenrm
A general technique is outlined for investigating
supersymmetry properties of a charged spin-$\half$ quantum particle
in time-varying electromagnetic fields.
The case of a time-varying uniform magnetic induction is examined
and shown to provide a physical realization
of a supersymmetric quantum-mechanical system.
Group-theoretic methods are used to factorize the relevant
Schr\"odinger equations and obtain eigensolutions.
The supercoherent states for this system are constructed.

}

\vfill
\newpage

\baselineskip=28pt

%***************************************************************************
%\include{m1}
\section{Introduction}

The quantum behavior of a nonrelativistic charged spin-$\half$ particle
in the presence of a constant and uniform magnetic induction
is of importance in many physical contexts.
The wave functions for this system are solutions of the Pauli equation,
which has two components corresponding to the two possible orientations
of the spin.
Each component has an energy spectrum consisting of a tower of equally
spaced levels,
called Landau levels
\cite{landau}.
The two sets of Landau levels are degenerate,
except for the ground state.
This system is known to provide a physical realization of
supersymmetric quantum mechanics
\cite{jackiw,hkn}.
The supersymmetry generators act to reverse the particle spin,
thereby mapping one tower of Landau levels into the other.

The classical motion of a point charge in a
constant and uniform magnetic induction is rotation
about a circle in the plane perpendicular to the magnetic field.
This motion is most closely reproduced in the quantum system
by coherent states
\cite{kszfg,m1,fk1},
for which the expectation values of the charge's coordinates
follow the classical cyclotron motion.
Coherent states can also be introduced
for the spin-$\half$ Landau system.
The presence of the supersymmetry makes possible
an extension of these states, resulting in supercoherent states
\cite{fknt1}.

It is natural to ask whether the notions of supersymmetry
and of supercoherent states can be introduced in the context of
the motion of a charged spin-$\half$ particle
in more general electromagnetic fields.
As the construction presented in Ref. \cite{fknt1}
relies on the factorization of the hamiltonian,
it is not apparent \it a priori \rm
how to handle more complicated situations.
The present paper addresses this issue.
We demonstrate that a group-theoretic analysis
can provide the key to a supersymmetric factorization.
Here,
we focus primarily on the case of a uniform but time-dependent
magnetic induction as an explicit example.
However, the formulation of the problem and the methods used
are applicable in a broader context.

We also use the results to obtain supercoherent states.
Our construction extends the previously developed
coherent states for a spinless charge in time-dependent
magnetic (and electric) fields
\cite{m3,m4,m5}.
This earlier approach
used time-dependent integrals of the motion satisfying
an oscillator algebra and the standard displacement-operator method
\cite{kla,gla,per}.

In Sec. 2,
we establish our notation and perform a first separation
of variables,
using the invariance of the Schr\"odinger operator
under translations along the direction of the magnetic induction.
A rotated variable set is introduced that simplifies much of
the subsequent analysis.
The group-theoretic analysis of the resulting equations
is presented in Sec. 3.
We seek symmetries of the problem using the methods detailed in
Ref. \cite{miller1} and applied in Refs. \cite{drt1,drt2,drt3}.
The complexified symmetry algebra is constructed in Sec. 3.2.
This generalizes the dynamical symmetry group
for the constant-induction case, presented in
Ref. \cite{m10}.

We use these results to develop several factorization schemes,
which are given in Sec. 4.
The solution to the relevant Schr\"odinger equation is
obtained in Sec. 5,
using group-theoretic techniques and some representation
theory taken from
Ref. \cite{drt4}.
In Sec. 6,
we extend these expressions to solutions of the Pauli equation,
and in Sec. 7 the supersymmetry is explicitly identified.
Finally,
we construct the relevant supercoherent states in Sec. 8,
thereby completing the generalization of the problem
treated in Ref. \cite{fknt1}.
The coherent states for a single tower of levels,
allowing for the time variation,
are contained as a limit of these supercoherent
states and are closely related to those constructed from integrals
of motion in
Ref. \cite{m9}.
The Appendix demonstrates the reduction of all these results
to the time-independent case.

\newpage

%***********************************************************************
%\include{m2}
\section{The Time-Dependent Landau Problem}

Consider a  nonrelativistic
spin-$\lfrac{1}{2}$ particle of mass $M$ and charge $e$
moving with momentum ${\bf p}$ in a time-dependent electromagnetic field
with four-vector potential $(\phi, {\bf A})$.
The Pauli equation for this system is
\begin{equation}
\{\frac{1}{2M}[{\bf \sigma}\cdot ({\bf p} - e{\bf A(r},t))]^2
 +
e\phi({\bf r},t)\}\Psi_3({\bf r},t) = i\partial_t\Psi_3({\bf r},t),
\label{e:a1}
\end{equation}
where $\Psi_3$ is a two-component wavefunction in three space dimensions
and the quantity ${\bf \sigma} = (\sigma _x, \sigma_y, \sigma_z)$
is a vector consisting of the three Pauli matrices.

In this paper,
we shall assume that the scalar potential $\phi({\bf r},t)$ is zero
and that the vector potential ${{\bf A(r},t)}$
describes a uniform, time-dependent magnetic induction $\bf B$.
For convenience, we work in in vacuo where the magnetic induction $\bf B$
is related to the magnetic field $\bf H$ by $\bf B = \mu_0 \bf H$,
and we choose the cylindrical gauge
\begin{equation}
A_x = -\lfrac{1}{2} By,~~A_y = \lfrac{1}{2} Bx. \label{e:a4}
\end{equation}
Writing the upper and lower components of $\Psi_3$ as
$\Psi_{3+}$ and $\Psi_{3-}$,
the Pauli equation reduces to the two equations
\begin{equation}
(T_x^2+T_y^2+T_z^2+2Me\phi\mp eB-2iM\partial _t)\Psi_{3\pm} = 0, \label{e:a5}
\end{equation}
where
\begin{equation}
T_x = p_x+eBy/2,~~T_y = p_y-eBx/2,~~T_z = p_z. \label{e:a6}
\end{equation}
Substituting for $p_x$, $p_y$, and $p_z$ the usual operator forms,
these equations can be rewritten as
\begin{equation} S_{3\pm}\Psi_{3\pm} = 0 ,\label{e:a7}
\end{equation}
where
\begin{equation} S_{3\pm} = \delt+iwL_z
-{\lfrac{1}{4}}w^2(x^2+y^2)\mp w + 2i\partial _{\tau} \label{e:a8}
\end{equation}
are called the Schr\"odinger operators in three space dimensions
and where we have set
\begin{equation}
\hbar = 1~~,~~~~\tau = t/M~~,~~~~eB(\tau) = w(\tau)~~,~~~~
L_z = y\partial _x -x\partial _y~~.\label{e:a10}
\end{equation}

The operators $S_{3\pm}$ commute with the $z$-translation operator $\prt_z$.
This implies \cite{miller1} that we can separate Eq. (\ref{e:a7})
with respect to $z$.
Set
\begin{equation}
\Psi_{3\pm}({\bf r},\tau) = \Psi_{\pm}(x,y,\tau)Z_{\kappa_{\pm}}(z)~~.
\label{e:a12}
\end{equation}
The functions $Z_{\kappa_{\pm}}(z)$ satisfy the eigenvalue problems
\begin{equation}
-i\partial_zZ_{\kappa_{\pm}} = \kappa_{\pm} Z_{\kappa_{\pm}}, \label{e:a13}
\end{equation}
with the usual plane wave solutions.
This procedure reduces Eq. (\ref{e:a7}) to
\begin{equation}
S_{2\pm}\Psi_{\pm} = 0~~, \label{e:a14}
\end{equation}
where the Schr\"odinger operators $S_{2\pm}$ in two space dimensions are
\begin{equation}
S_{2\pm}= \delw +iwL_z + 2i\partial_{\tau} -2h_2(x^2+y^2) - 2h_{0\pm},
\label{e:a15}
\end{equation}
with $2h_{0\pm}=\mp w+\kappa_{\pm}^2$ and $2h_2 = w^2/4$.
These new equations depend only on the variables $x$, $y$, and $\tau$.
The quantities $\kappa_{\pm}$ are the constants of separation.

It is convenient for our analysis to eliminate the $iwL_z$ term from
(\ref{e:a15}).
Introduce the operator
\begin{equation}
R = exp[\eta(\tau)L_z], \label{e:a17}
\end{equation}
where $\eta(\tau)$ is to be chosen below
to eliminate $iw(\tau)L_z$ from the expressions for $S_{2\pm}$.
Define rotated solutions $\Theta_{\pm}$ by
\begin{equation}
\Psi_{\pm} = R^{-1}\Theta_{\pm}. \label{e:a18}
\end{equation}
The rotated Schr\"odinger operators $\S_{2\pm}$ are given by
\bea
\S_{2\pm} & = & RS_{2\pm}R^{-1} = exp[\eta L_z]S_{2\pm}exp[-\eta L_z] \nonumber
\\
& = & \delw + iw(\tau)L_z +2i\partial_{\tau} -2i\dot{\eta}L_z - 2h_2(x^2+y^2) -
2h_{0\pm}.\label{e:a22}
\eea
Setting
\begin{equation}
\eta(\tau) = \lfrac{1}{2}\int_{\tau}w(\mu)d\mu \label{e:a23}
\end{equation}
simplifies the expressions to
\begin{equation}
\S_{2\pm}= \delw +2i\partial_{\tau} - 2h_2(x^2+y^2)-2h_{0\pm}.
 \label{e:a24}
\end{equation}
These operators are the Schr\"odinger operators for time-dependent isotropic
harmonic oscillators.
In the rotated frame,
Eq. (\ref{e:a14}) becomes
\begin{equation}
\S_{2\pm}\Theta_{\pm} = 0 .\label{e:a25}
\end{equation}

Throughout the remainder of this paper,
we use
the usual roman letters to represent operators in the
original space of the problem (\ref{e:a14}), and
script letters to represent operators associated with the rotated equations
(\ref{e:a25}).

\newpage

%************************************************************************
%\include{m3}
\section{Lie Symmetries}

If $w=eB$ is time-independent,
the solution of Eq.\ (\ref{e:a25}) is possible by a direct treatment.
In section IV of \cite{fknt1}, the separation of variables was performed
by setting $p_z = 0$ and $\Psi_{\pm} (x,y,\tau) = \psi_{\pm}
(x,y)e^{-iE_{\pm}\tau}$,
and the resulting differential equation was factored into raising and lowering
operators.
However, when $w$ varies with time it is no longer immediately apparent how to
separate
variables or how to identify appropriate raising and lowering operators into
which the
the differential equation (\ref{e:a14}) can be factored.
Instead, we proceed with a systematic approach that makes use of
symmetries of the Schr\"odinger operators $\S_{2\pm}$ in Eq. (\ref{e:a24}).

To simplify the expressions in this section and in Secs. 4 and 5,
we write equations only for the case associated with
the Schr\"odinger operator $\S_{2+}$,
rather than both $\S_{2\pm}$ at once.
We also denote $\S_{2+}$ by $\S_2$,
and $h_{0+}$ by $h_0$.
Analogous expressions for the problem with
$\S_{2-}$ can be found by replacing
all occurrences of $h_{0}=h_{0+}$ by $h_{0-}$.
This means that all the results obtained here and in the
subsequent two sections have duplicate forms.
We discuss the role and significance of this duality beginning in Sec. 6.

\subsection{The Symmetry Operators}

The symmetries we seek for the Schr\"odinger operator
$\S_{2}$
have the form \cite{miller1,drt1}
\begin{equation}
\tsL = \A(x,y,\tau)\partial_{\tau} + \Bo(x,y,\tau)\partial_x +
\Bw(x,y,\tau)\partial_y + C(x,y,\tau).\label{e:b1}
\end{equation}
These operators generate space-time transformations.
For these space-time transformations to be symmetries of (\ref{e:a24}),
they must satisfy the commutator \cite{miller1,drt1}
relation
\begin{equation}
[\S_2,\tsL] = \Lambda(x,y,\tau)\S_2,\label{e:b2}
\end{equation}
where $\Lambda(x,y,\tau)$ is some function of the space-time variables.
In the unrotated space of Eq. (\ref{e:a14}),
this equation takes the form
\begin{equation}
[S_2,\tL] = \lambda(x,y,\tau)S_2,\label{e:b4}
\end{equation}
where we have defined
\begin{equation}
\tL =  R^{-1}\tsL R =
\tilde{A}(x,y,\tau)\partial_{\tau}+\tilde{B}^1(x,y,\tau)\partial_x
+\tilde{B}^2(x,y,\tau)\partial_y+\tilde{C}(x,y,\tau)\label{e:b5}
\end{equation}
and
\begin{equation}
\lambda(x,y,\tau) = R^{-1}\Lambda(x,y,\tau)R. \label{e:b6}
\end{equation}
We next proceed to establish and solve a set of differential equations that
determine the explicit form of Eq. (\ref{e:b1}).

Substituting (\ref{e:a24}) for
$\S_{2}$ and (\ref{e:b1}) for $\tsL$ in Eq. (\ref{e:b2}),
we obtain a system of partial differential equations for the coefficients
$\A$, $\Bo$, $\Bw$, and $\C$:
\begin{eqnarray}
 & \A_x = \A_y = 0~~,~~~~ 2\Bo_x = 2\Bw_y = \Lambda~~,~~~~ \Bo_y + \Bw_x = 0, &
 %\label{e:b9} \\*[1.5mm]
 \nonumber \\*[1.5mm]
 & \A_{xx} + \A_{yy} + 2i\A_{\tau} = 2i\Lambda , &
 %\label{e:b10} \\*[1.5mm]
 \nonumber \\*[1.5mm]
 & \Bo_{xx} + \Bo_{yy} + 2i\Bo_{\tau} + 2C_x = 0~~,~~~~
 \Bw_{xx} + \Bw_{yy} + 2i\Bw_{\tau} + 2C_y = 0. & \label{e:b12}
\end{eqnarray}
We solve (\ref{e:b12}) in the usual manner \cite{drt1} to obtain
\begin{eqnarray}
 & \A = \A(\tau)~~,~~~~
  \Lambda = \dot{\A}, &
 %\label{e:b14} \\*[1.5mm]
 \nonumber \\*[1.5mm]
 & \Bo = \lfrac{1}{2}\dot{\A}x + \beta^1_2y + \epsilon^1(\tau)~~,~~~~
 \Bw = -\beta^1_2x + \lfrac{1}{2}\dot{\A}y + \epsilon^2(\tau), &
 %\label{e:b16} \\*[1.5mm]
 \nonumber \\*[1.5mm]
 & \C = -\lfrac{i}{4}\ddot{\A}(x^2+y^2) - i\dot{\epsilon}^1x -
i\dot{\epsilon}^2y
 + \epsilon(\tau), & \label{e:b17}
\end{eqnarray}
where the $\tau$-dependent coefficients $\A$, $\epsilon^1$, $\epsilon^2$,
and $\epsilon$ satisfy:
\begin{eqnarray}
 & \tdot{\A} + 8h_2\dot{\A} + 4\dot{h}_2\A = 0, & \label{e:b18} \\*[1.5mm]
 & \ddot{\epsilon}^1+2h_2\epsilon^1 = 0, & \label{e:b19} \\*[1.5mm]
 & \ddot{\epsilon}^2+2h_2\epsilon^2 = 0, & \label{e:b20} \\*[1.5mm]
 & i\dot{\epsilon} - \lfrac{i}{2}\ddot{\A} + \dot{k}_0\A + h_0\dot{\A} = 0. &
\label{e:b21}
\end{eqnarray}

Since Eqs. (\ref{e:b19}) and (\ref{e:b20}) have the same form, they are
satisfied by particular solutions $\chi_1(\tau)$ and $\chi_2(\tau)$
that also have the same form.
The coefficient of $\dot{\epsilon}_1$ is zero in (\ref{e:b19}),
so the wronskian $W(\chi_1,\chi_2)$ must be constant \cite{drt1}.
We choose to scale the two solutions so that
\begin{equation}
W(\chi_1,\chi_2) = \chi_1\dot{\chi}_2-\dot{\chi}_1\chi_2 = 1. \label{e:b22}
\end{equation}
The general solutions then have the form
\begin{eqnarray}
\epsilon^1(\tau) & = & \beta^{11}\chi_1(\tau)+\beta^{12}\chi_2(\tau),
 \nonumber \\*[1.5mm]
 %\label{e:b23} \\*[1.5mm]
\epsilon^2(\tau) & = & \beta^{21}\chi_1(\tau)+\beta^{22}\chi_2(\tau).
 \label{e:b24}
\end{eqnarray}
It is known \cite{drt1} that if $\chi_1$ and $\chi_2$ are solutions of
(\ref{e:b19}), then
\begin{equation}
\pt_1(\tau) = (\chi_1)^2,~~\pt_2(\tau) =
(\chi_2)^2,~~\pt_3(\tau)=2\chi_1\chi_2,
\label{e:b25}
\end{equation}
are particular solutions of (\ref{e:b18}).
The general solution is the linear combination
\begin{equation}
\A(\tau) = \beta^1\{\pt_1(\tau)\} + \beta^2\{\pt_2(\tau)\} +
 \beta^3\{\pt_3(\tau)\},
\label{e:b26}
\end{equation}
where $\beta^1$, $\beta^2$, and $\beta^3$ are real constants.

At this stage, Eq. (\ref{e:b21}) can be integrated to yield
\begin{equation}
\epsilon(\tau) = \beta^1\{\lfrac{1}{2}\dot{\pt}_1+ik_0\pt_1\} +
 \beta^2\{\lfrac{1}{2}\dot{\pt}_2+ik_0\pt_2\} + \beta^3\{\lfrac{1}{2}
 \dot{\pt}_3+ik_0\pt_3\} + \beta^4\{i\}. \label{e:b27}
\end{equation}
The remaining coefficients in the Lie derivative are then found to be
\begin{eqnarray}
\Bo & = & \beta^1\{\lfrac{1}{2}\dot{\pt}_1x\} +
 \beta^2\{\lfrac{1}{2}\dot{\pt}_2x\} + \beta^3\{\lfrac{1}{2}\dot{\pt}_3x\} +
 \beta^1_2\{ y\} + \beta^{11}\{\chi_1\} + \beta^{21}\{\chi_2\},
 %\label{e:b28} \\*[1.5mm]
 \nonumber \\*[1.5mm]
\Bw & = & - \beta^1_2\{x\} + \beta^1\{\lfrac{1}{2}\dot{\pt}_1y\} +
 \beta^2\{\lfrac{1}{2}\dot{\pt}_2y\} + \beta^3\{\lfrac{1}{2}\dot{\pt} _3y\} +
 \beta^{21}\{\chi_1\} + \beta^{22}\{\chi_2\},
 %\label{e:b29} \\*[1.5mm]
 \nonumber \\*[1.5mm]
\C  & = & \beta^1\{-\lfrac{i}{4}\ddot{\pt}_1(x^2+y^2) + \lfrac{1}{2}\dot{\pt}_1
 +ih_0\pt_1\} + \beta^2\{-\lfrac{i}{4}\ddot{\pt}_2(x^2+y^2) +
 \lfrac{1}{2}\dot{\pt}_2 +ih_0\pt_2\} \nonumber \\
    &   & ~~~~~~~~ + \beta^3\{-\lfrac{i}{4}\ddot{\pt}_3(x^2+y^2) +
 \lfrac{1}{2}\dot{\pt}_3 +ih_0\pt_3\}
 + \beta^{11}\{-ix\dot{\chi}_1\} + \beta^{12}\{-ix\dot{\chi}_2\}
 \nonumber \\ &   & ~~~~~~~~
 + \beta^{21}\{-iy\dot{\chi}_1\}
 + \beta^{22}\{-iy\dot{\chi}_2\} + \beta^4\{i\}. \label{e:b30}
\end{eqnarray}

Finally, the generators of the symmetry group of the Schr\"odinger operators
$\S_{2}$ can be obtained, by direct substitution.
Three of them  have the form
\begin{equation}
\tsL_j = \pt_j\partial_{\tau}+\lfrac{1}{2}\dot{\pt}_j(x\partial_x+y\partial_y)
-
\lfrac{i}{4}\ddot{\pt}_j(x^2+y^2) + \lfrac{1}{2}\dot{\pt}_j + ih_0\pt_j,
 \label{e:b31}
\end{equation}
for $j=1,2,3$.
These operators satisfy the commutation relations of an
$sl(2,{\bf R})$ Lie algebra \cite{drt1}:
\begin{equation} [\tsL_3,\tsL_1] = -2\tsL_1,~~[\tsL_3,\tsL_2] =
2\tsL_2,~~[\tsL_1,\tsL_2] = \tsL_3. \label{e:b32} \end{equation}
Another generator is $L_z$, which spans an $o(2)$ algebra and commutes with
the $\tsL_j$:
\begin{equation}
[L_z,\tsL_j] = 0. ~~~j=1,2,3. \label{e:b34}
\end{equation}
The remaining five generators span a Heisenberg-Weyl algebra, $w_2$, in
two-dimensional space \cite{drt1}.  These operators have the forms
\begin{eqnarray}
 & E  = i, &
 %\label{e:b35} \\*[1.5mm]
 \nonumber \\*[1.5mm]
 & \tsJ_1 = \chi_1\partial_x - ix\dot{\chi}_1,~~\tsJ_2 = \chi_2\partial_x -
 ix\dot{\chi}_2, &
 %\label{e:b36}\\*[1.5mm]
 \nonumber \\*[1.5mm]
 & \tsK_1 = \chi_1\partial_y - iy\dot{\chi}_1,~~\tsK_2 = \chi_2\partial_y -
 iy\dot{\chi}_2. & \label{e:b37}
\end{eqnarray}
Their nonzero commutation relations are
\begin{equation}
[\tsJ_1,\tsJ_2] = -\E,~~~[\tsK_1,\tsK_2] = -\E. \label{e:b38}
\end{equation}

The full symmetry algebra is the Schr\"odinger algebra in two space dimensions,
$(sl(2,{\bf R})\oplus o(2))\diamond w_2$.
The remaining commutation relations are
\begin{eqnarray}
 & [\tsL_1,\tsJ_1] = 0,~~[\tsL_2,\tsJ_1] = -\tsJ_2,~~[\tsL_3,\tsJ_1] = -\tsJ_1,
 & \nonumber \\*[1.5mm]
 & [\tsL_1,\tsJ_2] = \tsJ_1,~~[\tsL_2,\tsJ_2] = 0,~~[\tsL_3,\tsJ_2] = \tsJ_2,
 & \nonumber \\*[1.5mm]
%& \label{e:b39} \\*[1.75mm]
 & [\tsL_1,\tsK_1] = 0,~~[\tsL_2,\tsK_1] = -\tsK_2,~~[\tsL_3,\tsK_1] = -\tsK_1,
 & \nonumber \\*[1.5mm]
 & [\tsL_1,\tsK_2] = \tsK_1,~~[\tsL_2,\tsK_2] = 0,~~[\tsL_3,\tsK_2] = \tsK_2,
 & \nonumber \\*[1.5mm]
%& \label{e:b40} \\*[1.75mm]
 & [L_z,\tsJ_{\alpha}] = \tsK_{\alpha},~~[L_z,\tsK_{\alpha}] = \tsJ_{\alpha},
 ~~\alpha = 1,2.
 & \label{e:b41} \end{eqnarray}

\subsection{Complexification of the Symmetry Algebra}

To work with Hermitian or Hermitian-conjugate operators, we need to complexify
the symmetry algebra \cite{drt2}.
We begin by looking at the solutions of the
differential equations (\ref{e:b19}) and  (\ref{e:b20}).
The real solutions to these equations were denoted $\chi_1(\tau)$ and
$\chi_2(\tau)$.
 We can obtain complex solutions from these by defining
\begin{equation}
\xi(\tau) = \lfrac{1}{\sqrt{2}}(\chi_1(\tau)+i\chi_2(\tau)),~~\bx(\tau) =
\lfrac{1}{\sqrt{2}}(\chi_1(\tau)-i\chi_2(\tau)).  \label{e:c2}
\end{equation}
In this case, the general solutions to
Eqs. (\ref{e:b19}) and (\ref{e:b20}) can be written
\begin{equation}
\epsilon^1(\tau) = \beta^{11}\{\xi\} + \beta^{12}\{\bx\},~~~\epsilon^2(\tau) =
\beta^{21}\{\xi\} + \beta^{22}\{\bx\}.\label{e:c3}
\end{equation}
The Wronskian of these solutions is
\begin{equation}
W(\xi,\bx) = \xi\dot{\bx}-\dot{\xi}\bx = -iW(\chi_1,\chi_2) =-i.
\label{e:c4} \end{equation}

Complex solutions to the differential equation (\ref{e:b18}) can be written in
analogy
to the real solutions (\ref{e:b25}).  We have
\begin{equation}
\varphi_1 = \xi^2,~~~\varphi_2 = \bx^2,~~~\varphi_3 = 2\xi\bx. \label{e:c5}
\end{equation}

Then, in terms of these complex solutions, three of the generators are
\begin{equation}
{\hat{\L}}_j = \varphi_j\partial_{\tau} +
\lfrac{1}{2}\dot{\varphi}_j(x\partial_x+y\partial_y)
 -\lfrac{i}{4}\ddot{\varphi}_j(x^2+y^2)
+ \lfrac{1}{2}\dot{\varphi}_j +ih_0\varphi_j, \label{e:c6}
\end{equation}
where $j=1,2,3$.  We can express these operators as linear combinations of the
original operators (\ref{e:b31}), which form the basis
 of the $sl(2,{\bf R})$ subalgebra, in the following way:
\beq
  {\hat{\L}}_1=\lfrac{1}{2}(\tsL_1-\tsL_2+i\tsL_3)~~,~~~~
  {\hat{\L}}_2=\lfrac{1}{2}(\tsL_1-\tsL_2-i\tsL_3)~~,~~~~
  {\hat{\L}}_3=\tsL_1+\tsL_2.  \label{e:c9}
\eeq
It is more convenient to define the operators
\begin{equation}
\M_3= i{\hat{\L}}_3,~~~\M_+={\hat{\L}}_2,~~~\M_-=-{\hat{\L}}_1.\label{e:c10}
\end{equation}
These operators satisfy the commutation relations
\begin{equation}
[\M_+,\M_-] = -\M_3,~~~[\M_3,\M_+] = 2\M_+,~~~[\M_3,\M_-]=-2\M_-.\label{e:c11}
\end{equation}
The operators (\ref{e:c10}) form a basis for an $su(1,1)$ algebra.

The generator $L_z$ spans an $o(2)$ algebra, as before.

Finally, there are the five generators of a Heisenberg-Weyl algebra $w_2$:
\begin{eqnarray}
 & \J_- = \xi\partial_x-ix\dot{\xi},~~\J_+ = -\bx\partial_x+ix\dot{\bx}, &
 \nonumber \\*[1.5mm]
 %\label{e:c12} \\*[1.5mm]
 & \K_- = \xi\partial_y-iy\dot{\xi},~~\K_+ = -\bx\partial_x+iy\dot{\bx}, &
 \nonumber \\*[1.5mm]
 %\label{e:c13} \\*[1.5mm]
 & I= 1. & \label{e:c14}
\end{eqnarray}
These operators satisfy the nonzero commutation relations
\begin{equation}
[\J_-,\J_+] = I,~~~[\K_-,\K_+] = I.\label{e:c15}
\end{equation}
The operators $\J_{\pm}$ and $\K_{\pm}$ can also be expressed in terms of the
operators
$\J_{\alpha}$ and $\K_{\alpha}$, $\alpha=1,2$, as follows:
\begin{eqnarray}
 & \J_- = \lfrac{1}{\sqrt{2}}(\tsJ_1+i\tsJ_2),~~~\J_+ =
 \lfrac{1}{\sqrt{2}}(-\tsJ_1+i\tsJ_2), &
 \nonumber \\*[1.5mm]
 %\label{e:c16} \\*[1.5mm]
 & \K_- = \lfrac{1}{\sqrt{2}}(\tsK_1+i\tsK_2),~~~\K_+ =
 \lfrac{1}{\sqrt{2}}(-\tsK_1+i\tsK_2). &
\label{e:c17}
\end{eqnarray}

The remaining commutation relations of the full Schr\"odinger algebra are
\begin{eqnarray}
 & [L_z,\M_3] = [L_z,M_+] = [L_z,M_-] = 0,
 & \nonumber \\*[1.5mm]
%& \label{e:c18} \\*[1.5mm]
 & [L_z,\J_-] = \K_-,~~~[L_z,\J_+] = \K_+,
 & \nonumber \\*[1.5mm]
%& \label{e:c19} \\*[1.5mm]
 & [L_z,\K_-] = -\J_-,~~~[L_z,\K_+] = -\J_+,
 & \nonumber \\*[1.5mm]
%& \label{e:c20} \\*[1.5mm]
 & [\M_-,\J_-] = 0,~~[\M_+,\J_-] = -i\J_+,~~ [\M_3,\J_-] = -\J_-,
 & \nonumber \\*[1.5mm]
%&\label{e:c21} \\[1.5mm]
 & [\M_-,\J_+] = -i\J_-,~~[\M_+,\J_+] = 0,~~ [\M_3,\J_+] = \J_+,
 & \nonumber \\*[1.5mm]
%& \label{e:c22} \\*[1.5mm]
 & [\M_-,\K_-] = 0,~~[\M_+,\K_-] = -i\K_+,~~ [\M_3,\K_-] = -\K_-,
 & \nonumber \\*[1.5mm]
%& \label{e:c23} \\[1.5mm]
 & [\M_-,\K_+] = -i\K_-,~~[\M_+,\K_+] = 0,~~ [\M_3,\K_+] = \K_+.
 & \label{e:c24}
 \end{eqnarray}

It is also useful to have the expressions for the operators and equations
in the original, unrotated frame.
The operators of the Schr\"odinger algebra for (\ref{e:a14}) can be obtained
from (\ref{e:b5}) and are given by
\begin{eqnarray}
 & j_- = R^{-1}\J_-R = \J_-\cos(\eta) - \K_-\sin(\eta),
 & \nonumber \\*[1.5mm]
 & j_+ = R^{-1}\J_+R = \J_+\cos(\eta) - \K_+\sin(\eta), & \nonumber \\*[1.5mm]
 & k_- = R^{-1}\K_-R = \K_-\cos(\eta) + \J_-\sin(\eta), & \nonumber \\*[1.5mm]
 & k_+ = R^{-1}\K_+R = \K_+\cos(\eta) + \J_+\sin(\eta), & \label{e:c25}
\end{eqnarray}
and
\begin{eqnarray}
m_3 & = & R^{-1}\M_3R = \M_3 + \lfrac{i}{2}\varphi_3wL_z \nonumber \\*[1.5mm]
    & = & i\{\varphi_3(\partial_{\tau} + \lfrac{1}{2}wL_z) +
 \lfrac{1}{2}\dot{\varphi}_3(x\partial_x+y\partial_y)
%\nonumber \\ &   & ~~~~~~~~
 -\lfrac{i}{4}\ddot{\varphi}_3(x^2+y^2) +
 \lfrac{1}{2}\dot{\varphi}_3 + ih_0\varphi_3\}, \nonumber \\*[1.5mm]
m_+ & = & R^{-1}\M_+R = \M_+ +\lfrac{1}{2}\varphi_2wL_z \nonumber \\*[1.5mm]
    & = & \varphi_2(\partial_{\tau} + \lfrac{1}{2}wL_z) +
 \lfrac{1}{2}\dot{\varphi}_2(x\partial_x+y\partial_y)
%\nonumber \\ &   & ~~~~~~~~
 -\lfrac{i}{4}\ddot{\varphi}_2(x^2+y^2) +
 \lfrac{1}{2}\dot{\varphi}_2 + ih_0\varphi_2, \nonumber \\*[1.5mm]
m_- & = & R^{-1}\M_-R = \M_- - \lfrac{1}{2}\varphi_1wL_z \nonumber \\*[1.5mm]
    & = & -\varphi_1(\partial_{\tau} + \lfrac{1}{2}wL_z) -
 \lfrac{1}{2}\dot{\varphi}_1(x\partial_x+y\partial_y)
%\nonumber \\ &   & ~~~~~~~~
 +\lfrac{i}{4}\ddot{\varphi}_1(x^2+y^2) -
 \lfrac{1}{2}\dot{\varphi}_1 - ih_0\varphi_1. \label{e:c26}
\end{eqnarray}
The commutation relations (\ref{e:c10}), (\ref{e:c15}) and
(\ref{e:c24}) are preserved by the transformations (\ref{e:c25}) and
(\ref{e:c26}).
% so we can write
%\begin{eqnarray}
% & [m_+,m_-] = - m_3,~~~[m_3,m_+] = 2m_+,~~~[m_3,m_-] = -2m_-, & \label{e:c27}
% \\*[1.5mm]
% & [j_-,j_+] = I,~~~~[[k_-,k_+] = I, & \label{e:c28} \\*[1.5mm]
% & [L_z,j_-] = k_-,~~~~[L_z,j_+] = k_+, & \nonumber \\*[1.5mm]
% & [L_z,k_-] = -j_-,~~~~[L_z,k_+] = -j_+, & \label{e:c29} \\*[1.5mm]
% & [m_-,j_-] = [m_-,k_-] = [m_+,j_+] = [m_+,k_+] = 0, & \nonumber \\*[1.5mm]
% & [m_-,j_+] = -ij_-,~~~~[m_-,k_+] = -ik_-, & \nonumber \\*[1.5mm]
% & [m_+,j_-] = -ij_+,~~~~[m_+,k_-] = -ik_+, & \nonumber \\*[1.5mm]
% & [m_3,j_{\pm}] = \pm j_{\pm},~~~~[m_3,k_{\pm}] = \pm k_{\pm}. &
%%\label{e:c30}
%\end{eqnarray}

A more convenient choice for a basis for
$w_2$ can be made.   Define the operators
\begin{eqnarray}
a_- & = & \lfrac{1}{\sqrt{2}}(j_-+ik_-) =
\lfrac{e^{i\eta}}{\sqrt{2}}(\J_-+i\K_-)
% \nonumber \\*[1.5mm]
%    & = &
=\frac{e^{i\eta}}{\sqrt{2}}[\xi(\partial_x+i\partial_y)-i(x+iy)\dot{\xi}],
 %\label{e:c31} \\*[1.75mm]
 \nonumber \\*[1.5mm]
%a_+ & = & \lfrac{1}{\sqrt{2}}(j_+-ik_+) =
% \frac{e^{-i\eta}}{\sqrt{2}}(\J_+-i\K_+),\nonumber \\*[1.5mm]
%    & = &
% \frac{e^{-i\eta}}{\sqrt{2}}[-\bx(\partial_x-i\partial_y)+i(x-iy)\dot{\bx}],
% \label{e:c32} \\*[1.75mm]
c_- & = & \lfrac{1}{\sqrt{2}}(j_--ik_-) =
 \frac{e^{-i\eta}}{\sqrt{2}}(\J_--i\K_-)
%\nonumber \\*[1.5mm]
%    & = &
=\frac{e^{-i\eta}}{\sqrt{2}}[\xi(\partial_x-i\partial_y)-i(x-iy)\dot{\xi}].
 \label{e:c33}
 % \\*[1.75mm]
%c_+ & = & \lfrac{1}{\sqrt{2}}(j_++ik_+) =
%%\frac{e^{i\eta}}{\sqrt{2}}(\J_++i\K_+),
% \nonumber
%\\*[1.5mm]
%    & = &
% \frac{e^{i\eta}}{\sqrt{2}}[-\bx(\partial_x+i\partial_y)+i(x+iy)\dot{\bx}].
%\label{e:c34}
\end{eqnarray}
%The generators $m_3$, $m_{\pm}$ remain unchanged.
and their conjugates $a_+$ and $c_+$.
These operators are related to the raising and lowering operators $a$,
$a^{\dagger}$,
$c$, and $c^{\dagger}$ introduced in Ref. \cite{fknt1}.
Appendix A contains an explicit demonstration of the connection between the two
sets
of operators for the specific case of a charged particle
moving in a constant magnetic induction.

The nonzero commutation relations for the complexified Schr\"odinger algebra
in this new basis are:
\begin{eqnarray}
 & [m_+,m_-] = -m_3,~~[m_3,m_+] = 2m_+,~~[m_3,m_-] = -2m_-, &
 \label{e:c27za}\\*[1.5mm]
 & [a_-,a_+] = I,~~[c_-,c_+] = I, & \label{e:c35} \\*[1.5mm]
 & [m_3,a_{\pm}] = \pm a_{\pm},~~[m_3,c_{\pm}] = \pm c_{\pm}, & \label{e:c36}
 \\*[1.5mm]
 & [m_-,a_+] = -ic_-, ~~[m_-,c_+] = -ia_-,~~
%& \nonumber \\*[1.5mm]
%&
[m_+,a_-] = -ic_+, ~~[m_+,c_-] = -ia_+, & \label{e:c37} \\*[1.5mm]
 & [\L_z,a_{\pm}] = \mp a_{\pm}, ~~[\L_z,c_{\pm}] = \pm c_{\pm}. &
\label{e:c38}
\end{eqnarray}

\newpage

%**********************************************************************
%\include{m4}
\section{Factorization of the Schr\"odinger Equation}

In this section,
we continue to work with the Schr\"odinger operator $\S_2 = \S_{2+}$
as noted at the beginning of the previous section.
Our next goal is to establish candidate factorizations for this operator.

It suffices to work with the subalgebra
$\G = \{m_3,a_{\pm},c_{\pm},I\}\diamond \{\L_z\}$
of the Schr\"odinger algebra.
If we denote the oscillator subalgebra generated by $\{m_3,
a_{\pm},c_{\pm},I\}$
as $os(2)$, then $\G = os(2)\diamond o(2)$,
where $\L_z = iL_z$ is the generator of $o(2)$.
The span of the oscillator subalgebra satisfies the nonzero commutation
relations
(\ref{e:c35}), (\ref{e:c36}), and (\ref{e:c38}).

The first step is to calculate the operators $a_+a_-$ and $c_+c_-$.
We find
\begin{eqnarray}
a_+a_- & = & \lfrac{1}{2}[-\lfrac{1}{2}\varphi_3(\partial_{xx}+\partial_{yy}) +
\lfrac{i}{2}\dot{\varphi}_3(x\partial_x+y\partial_y) -
i(y\partial_x-x\partial_y)
\nonumber \\ &   & ~~~~~~~~~~~~
+ \dot{\xi}\dot{\bx}(x^2+y^2) +
 \lfrac{i}{2}\dot{\varphi_3} - 1]
\nonumber \\*[1.5mm]
       & = & \lfrac{1}{2}[-\lfrac{1}{2}\varphi_3\S_2 + \M_3 - \L_z -1]
% \label{e:d2} \\*[1.5mm] & = &
      =\lfrac{1}{2}[-\lfrac{1}{2}\varphi_3 S_2 + m_3 - \L_z -1],
 \label{e:d3}
\end{eqnarray}
and
\begin{eqnarray}
c_+c_- & = & \lfrac{1}{2}[-\lfrac{1}{2}\varphi_3(\partial_{xx}+\partial_{yy}) +
\lfrac{i}{2}\dot{\varphi}_3(x\partial_x+y\partial_y) +
i(y\partial_x-x\partial_y)
\nonumber \\ &   & ~~~~~~~~~~~~
+ \dot{\xi}\dot{\bx}(x^2+y^2) +
 \lfrac{i}{2}\dot{\varphi_3} - 1]
\nonumber \\*[1.5mm]
       & = & \lfrac{1}{2}[-\lfrac{1}{2}\varphi_3\S_2 + \M_3 + \L_z -1]
%\label{e:d5} \\*[1.5mm] & = &
     =\lfrac{1}{2}[-\lfrac{1}{2}\varphi_3 S_2 + m_3 + \L_z -1].
 \label{e:d6}
\end{eqnarray}

Rearranging Eqs. (\ref{e:d3}) and (\ref{e:d6}), we
obtain
\begin{eqnarray}
-\lfrac{1}{2}\pt_3\S_2 & = & 2a_+a_- - \M_3 + \L_z +1
= 2c_+c_- - \M_3 - \L_z +1, \label{e:d9}\\*[1.5mm]
-\lfrac{1}{2}\pt_3S_2 & = & 2a_+a_- - m_3 + \L_z +1
= 2c_+c_- - m_3 - \L_z +1, \label{e:d10}
\end{eqnarray}
where
\beq
\M_3 = i\{\varphi_3\partial_{\tau} +
 \lfrac{1}{2}\dot{\varphi}_3(x\partial_x+y\partial_y)
    -\lfrac{i}{4}\ddot{\varphi}_3(x^2+y^2) +
 \lfrac{1}{2}\dot{\varphi}_3 + ih_0\varphi_3\},\label{e:d11}
\eeq
and $m_3$ is given by (\ref{e:c26}).
Equation (\ref{e:d9}) provides two different factorization
schemes for the
Schr\"odinger operator for the two-dimensional time-dependent harmonic
oscillator,
while Eq. (\ref{e:d10}) represents two factorization schemes for the
Schr\"odinger operator for an electron in a time-dependent uniform magnetic
induction.
The equations are related to one another by
the rotation $R$ of (\ref{e:a17}).

We remind the reader again that
similar factorizations can be obtained for the Schr\"odinger
operators $S_{2-}$ and $\S_{2-}$.
Note also that the operators $a_{\pm}$, $c_{\pm}$ and $L_z$
are unaffected by the replacement of
$h_{0+}$ by $h_{0-}$.

\newpage

%*******************************************************************
%\include{m5}
\section{Solution of the Time-dependent Schr\"odinger Equation}

In this section,
we present solutions of the two-dimensional Schr\"odinger equations
obtained in Sec. 2.
The results are found by group-theoretic techniques,
and as such can be extended to more general situations than
the one considered here.
Once again,
we remind the reader that,
to minimize notational confusion,
results are presented only for the operator
$S_2 = S_{2+}$, the wavefunction $\Psi = \Psi_+$,
and rotated forms.
The analogous expressions for
$\S_{2-}$ and $\Psi_-$ can be found by replacing
$h_0 = h_{0+}$ with $h_{0-}$.

\subsection{Further Separation}

Denote the solution space of the Schr\"odinger equation
$\S_2\Psi = 0$ by $Q_{\S_2}$.
The generators of space-time transformations
for the Schr\"odinger equation, say $\tL$ of (\ref{e:b5}),
must satisfy Eq. (\ref{e:b4}).
Then, for $\Psi \in Q_{\S_2}$, we see that
\begin{equation}
[\S_2,\tL]\Psi = \lambda \S_2\Psi = 0.\label{e:e1}
\end{equation}
This means the generators are constants of the motion \cite{messiah1}.
In particular,
the generators $\L_z$ and $m_3$ are two commuting constants of the motion.
Therefore,
we can find a set of common eigenfunctions, $\Psi_{u,l}$,
labeled by the eigenvalues $u$ and $l$ of $m_3$ and $\L_z$, respectively:
\begin{equation}
m_3\Psi_{u,l} = u\Psi_{u,l},~~~\L_z\Psi_{u,l}=l\Psi_{u,l}.\label{e:e2}
\end{equation}
Recall from (\ref{e:a18}) that $\Psi_{u,l} = e^{-\eta L_z}\Theta_{u,l}$.
Hence, we have
\begin{equation}
\M_3\Theta_{u,l} = e^{\eta L_z}m_3 e^{-\eta L_z}\Theta_{u,l} =
u\Theta_{u,l}.\label{e:e3}
\end{equation}
Similarly, we obtain
\begin{equation}
\L_z\Theta_{u,l} = l\Theta_{u,l} \label{e:e4}
\end{equation}
for the second eigenvalue equation.

In Eq. (\ref{e:e2}), substitute the operator (\ref{e:d11}),
where $\pt_3$ is given by (\ref{e:c5}).
The resulting equation is a first-order partial
differential equation for $\Theta_{u,l}$, which can be solved by the method of
characteristics. The general solution has the form
\begin{equation}
\Theta_{u,l} =
 \exp^{\{i\R(\zeta_1,\zeta_2,\mu)\}}\psi_{u,l}(\zeta_1,\zeta_2)T_u(\mu),
\label{e:e7}
\end{equation}
where the separable coordinates are
\begin{equation}
\zeta_1 = \frac{x}{\varphi_3^{1/2}},~~~\zeta_2
 =\frac{y}{\varphi_3^{1/2}},~~~\mu=\tau,
\label{e:e8}
\end{equation}
and where
\begin{eqnarray}
 & \R =  \lfrac{1}{4}\dv_3(\zeta_1^2+\zeta_2^2), & \nonumber \\*[1.5mm]
 & T_u(\mu) = \varphi_3^{-1/2}\bigl (\frac{\bx}{\xi}\bigr )^{u/2}\exp{(-iK_0)},
&
\nonumber \\*[1.5mm]
 & \int_{\mu}ds~h_0 = K_0. & \label{e:e11}
\end{eqnarray}
For details of the integration, see the Appendix in Ref. \cite{drt2}.
Note that the function $\R$ cannot be written as a sum $\R_1(\zeta_1)
+ \R_2(\zeta_2) + \R_3(\mu)$ of arbitrary functions.
The solutions $\Theta_{u,l}$
are called $R$-separable solutions \cite{miller1} of the equation
$\S_2\Theta_{u,l}=0$.

Similarly, the functions $\Psi_{u,l}$ that solve the first eigenvalue problem
in (\ref{e:e2}) also solve the Schr\"odinger equation $S_2\Psi=0$.
Since $e^{-\eta L_z}$ commutes with $e^{i\R}$,
we obtain the solution to (\ref{e:a14}):
\begin{equation}
\Psi_{u,l} =  e^{i\R}e^{-\eta
 L_z}\psi_{u,l}(\zeta_1,\zeta_2)T_u(\mu),\label{e:e14}
\end{equation}
where $\R$ and $T_u$ are given by (\ref{e:e11}).
Substituting for $\Psi_{u,l}$ in (\ref{e:a14}) and suppressing the $u$ and $l$
labels, we obtain
\begin{equation}
\psi_{\zeta_1\zeta_1}+\psi_{\zeta_2\zeta_2}-(\zeta_1^2+\zeta_2^2)\psi + 2q\psi
=
 0,
\label{e:e15}
\end{equation}
which is the eigenvalue problem for a two-dimensional harmonic oscillator.

A further separation of variables can now be made.
We solve the eigenvalue equation
 (\ref{e:e4}), where
$\L_z$ is given by
\begin{equation}
 \L_z = i(y\partial_x-x\partial_y) =  i(\zeta_2\partial_{\zeta_1} -
\zeta_1\partial_{\zeta_2}),  \label{e:e16}
\end{equation}
and $\Theta_{u,l}$ by (\ref{e:e7}).  Since $\L_z$ commutes with
 $\R$, we obtain the eigenvalue problem
\begin{equation}
\L_z\psi_{u,l} = l\psi_{u,l},  \label{e:e18}
\end{equation}
which yields directly the first-order partial differential equation
\begin{equation}
\zeta_2{{\partial\psi}\over{\partial\zeta_1}} -
\zeta_1{{\partial\psi}\over{\partial\zeta_2}} = -il\psi.\label{e:e19}
\end{equation}
Solving this equation by the method of characteristics, we get the solution
\begin{equation}
\psi_{u,l} = \Upsilon_{u,l}(\rho)e^{il\theta},\label{e:e20}
\end{equation}
where the variables of separation $\rho$ and $\theta$ form a polar coordinate
system with
\begin{equation}
\rho^2 = \zeta_1^2+\zeta_2^2,~~~\theta =
 \tan^{-1}(\frac{\zeta_2}{\zeta_1}),\label{e:e21}
\end{equation}
and
\begin{equation}
\zeta_1 = \rho\cos{\theta},~~~~\zeta_2 = \rho\sin{\theta}. \label{e:e22}
\end{equation}
The eigenfunction $\Theta_{u,l}$ has the form
\begin{equation}
\Theta_{u,l} (\rho,\theta,\mu) = e^{i\R}\Upsilon(\rho)e^{il\theta}T_{u}(\mu),
\label{e:e22za}
\end{equation}
where $T_u(\mu)$ is given by (\ref{e:e11}) and
\begin{equation}
\R = \lfrac{1}{4}\dv_3\rho^2.\label{e:e22zb}
\end{equation}
The wave function $\Psi_{u,l} \in Q_{\S_2}$ has the form
\begin{eqnarray}
\Psi_{u,l} & = & e^{i\R}e^{i\eta\L_z}\Upsilon_{u,l}(\rho)e^{il\theta}T_{u}(\mu)
\nonumber\\*[1.5mm]
     & = &
 e^{i\R}\Upsilon_{u,l}(\rho)e^{il(\theta+\eta)}T_{u}(\mu),\label{e:e24}
\end{eqnarray}
and $T_{u}$ is given by (\ref{e:e11}).

\subsection{Rotation of the Operators}

To obtain the wave function $\Upsilon_{u,l}(\rho)$ using the ladder
operators $a_{\pm}$ and $c_{\pm}$, we transform the
generators of the $os(2)$ subalgebra into a form that can be written as a
product of a time-dependent function and an operator depending only on
$\rho$ and $\theta$.  Let $o$ be some operator which acts on the
manifold of solutions $Q_{S_2}$.  Then, we have
\begin{eqnarray}
o\Psi & = & oe^{i\R}e^{-\eta L_z}\psi(\zeta_1,\zeta_2)T(\mu) \nonumber
 \\*[1.5mm]
      & = & e^{i\R}e^{-\eta L_z}e^{\eta L_z} e^{-i\R}oe^{i\R}e^{-\eta
L_z}\psi(\zeta_1,\zeta_2)T(\mu) \nonumber \\*[1.5mm]
      & = & e^{i\R}e^{-\eta L_z}O\psi(\zeta_1,\zeta_2)T(\mu), \label{e:e26}
\end{eqnarray}
where the new operator $O$ is defined to be
\begin{equation}
O = e^{\eta L_z} e^{-i\R}oe^{i\R}e^{-\eta L_z}, \label{e:e27}
\end{equation}
with $\R$ given by (\ref{e:e11}) or (\ref{e:e22zb}).  The operator $O$ acts on
the product space of functions denoted by
$\{\Upsilon_{u,l}(\rho)e^{il\theta}T_u(\mu)\}$.  Note that the generator
$\L_z$ of the $o(2)$ algebra is invariant under this transformation.

Let the operator $o$ be a generator of the Heisenberg-Weyl algebra $w_2$.
For $a_{\pm}$, given by (\ref{e:c33}), we get
\begin{eqnarray}
e^{\eta L_z} e^{-i\R}a_-e^{i\R}e^{-\eta L_z} & = & \lfrac{1}{2}\bigl (
\frac{\bx}{\xi} \bigr )^{-\frac{1}{2}}e^{i\theta}
[\partial_{\rho}+\rho+\frac{i}{\rho}\partial_{\theta}] \nonumber\\*[1.5mm]
                                             & = & A_-, \label{e:e31}
\end{eqnarray}
and
\begin{eqnarray}
e^{\eta L_z} e^{-i\R}a_+e^{i\R}e^{-\eta L_z} & = & \lfrac{1}{2}\bigl (
\frac{\bx}{\xi} \bigr )^{\frac{1}{2}}e^{-i\theta}
[-\partial_{\rho}+\rho+\frac{i}{\rho}\partial_{\theta}]\nonumber \\*[1.5mm]
                                             & = & A_+. \label{e:e32}
\end{eqnarray}
For the remaining operators $c_{\pm}$ of the $w_2$ subalgebra, we obtain
\begin{eqnarray}
e^{\eta L_z} e^{-i\R}c_-e^{i\R}e^{-\eta L_z} & = & \lfrac{1}{2}\bigl
(\frac{\bx}{\xi} \bigr )^{-\frac{1}{2}}e^{-i\theta}
[\partial_{\rho}+\rho-\frac{i}{\rho}\partial_{\theta}]\nonumber \\*[1.5mm]
                                             & = & C_-, \label{e:e33}
\end{eqnarray}
and
\begin{eqnarray}
e^{\eta L_z} e^{-i\R}c_+e^{i\R}e^{-\eta L_z} & = & \lfrac{1}{2}\bigl
(\frac{\bx}{\xi} \bigr )^{\frac{1}{2}}e^{i\theta}
[\partial_{\rho}+\rho-\frac{i}{\rho}\partial _{\theta}]\nonumber \\*[1.5mm]
                                             & = & C_+. \label{e:e34}
\end{eqnarray}
In later applications,
it is also useful to introduce the operators
\begin{eqnarray}
a & = & \lfrac{1}{2}e^{i\theta}[\partial_{\rho}+\rho+
\lfrac{i}{\rho}\partial_{\theta}]~~,~~~~
\adg = \lfrac{1}{2}e^{-i\theta}[-\partial_{\rho}+\rho+
\lfrac{i}{\rho}\partial_{\theta}],\nonumber\\*[1.75mm]
c & = & \lfrac{1}{2}e^{-i\theta}[\partial_{\rho}+\rho-
\lfrac{i}{\rho}\partial_{\theta}]~~,~~~~
\cdg = \lfrac{1}{2}e^{i\theta}[\partial_{\rho}+\rho-
\lfrac{i}{\rho}\partial_{\theta}],\label{e:e57}
\end{eqnarray}
where
\begin{eqnarray}
 & A_- =  \bigl ( \frac{\bx}{\xi}\bigr )^{-\frac{1}{2}}a,~~~A_+  =
\bigl ( \lfrac{\bx}{\xi}\bigr )^{\frac{1}{2}}\adg, & \nonumber \\*[2.0mm]
 & C_-  = \bigl ( \frac{\bx}{\xi}\bigr )^{-\frac{1}{2}}c,~~~C_+ =
\bigl ( \frac{\bx}{\xi}\bigr )^{\frac{1}{2}}\cdg. & \label{e:e59}
\end{eqnarray}
The operators $a$, $\adg$, $c$, and $\cdg$ are the analogues of the
corresponding operators in \cite{fknt1}.
See the Appendix.

Finally, the generator $m_3$ transforms as
\begin{eqnarray}
e^{\eta L_z}e^{-i\R}m_3e^{i\R}e^{-\eta L_z} & = & i\{\varphi_3\partial_{\mu} +
\lfrac{1}{2}\dv_3 + ih_0\varphi_3\}. \nonumber \\*[1.5 mm]
                                            & = & M_3. \label{e:e35}
\end{eqnarray}
Note that all the spatial dependence has been removed in $M_3$.  It is now a
purely $\mu$-dependent operator.

\subsection{Reduction to a Radial Equation}

To complete the solution of Eqs. (\ref{e:a14}),
we shall exploit the representation theory of the Lie algebra
$\G = os(2)\diamond o(2)$.  It is more convenient to work in the
new basis
\begin{equation}
\{d_3,f_3,a_{\pm},c_{\pm},I\}\label{e:e35za}
\end{equation}
where the operators $d_3$ and $f_3$ are defined to be
\begin{equation}
d_3 = \lfrac{1}{2}(m_3+\L_z),~~~f_3 = \lfrac{1}{2}(m_3 - \L_z).\label{e:e36}
\end{equation}
The operators in this algebra satisfy the nonzero commutation relations:
\begin{eqnarray}
 & [m_3,a_{\pm}] = \pm a_{\pm},~~~~[a_-,a_+] = I, & \nonumber \\*[1.5mm]
 & [m_3,c_{\pm}] = \pm c_{\pm},~~~~[c_-,c_+] = I, & \nonumber \\*[1.5mm]
 & [\L_z,a_{\pm}] = \mp a_{\pm},~~~~[\L_z,c_{\pm}] = \pm c_{\pm}. &
\label{e:apxbc}
\end{eqnarray}

The detailed representation theory for this Lie algebra is presented
elsewhere \cite{drt4}.  The specific representation that is of interest
to us is denoted $\uparrow_{-\frac{1}{2}}\otimes \uparrow_{-\frac{1}{2}}$.
The \rep\ space is spanned by a set of vectors $\{\ketw{n}{m},~n,m\in \bZp\}$,
where $\bZp$ is the set of nonnegative integers.  The base for the Lie
algebra $\G$ acts on this space in the following way:
\begin{eqnarray}
 & f_3\ketw{n}{m} = (\half +n)\ketw{n}{m},~~d_3\ketw{n}{m} = (\half
+m)\ketw{n}{m},
\nonumber\\*[1.5mm]
 & a_+\ketw{n}{m} =
\sqrt{n+1}\ketw{n+1}{m},~~a_-\ketw{n}{m}=\sqrt{n}\ketw{n-1}{m},
\nonumber\\*[1.5mm]
 & c_+\ketw{n}{m} =
\sqrt{m+1}\ketw{n}{m+1},~~c_-\ketw{n}{m}=\sqrt{m}\ketw{n}{m-1},
\nonumber\\*[1.5mm]
 & I\ketw{n}{m} = \ketw{n}{m}.\label{e:apxbfea}
\end{eqnarray}
The \rep\ space is a normed space and we have
\begin{equation}
\langle n',m'\vert n,m\rangle = \delta_{n'n}\delta_{m'm}.\label{e:e40zz}
\end{equation}
We identify the solution spaces $Q_{S_2}$ and $Q_{\S_2}$ with the
unrotated and rotated \rep\ spaces
of the \irrep\ $\uparrow_{-\frac{1}{2}}\otimes \uparrow_{-\frac{1}{2}}$.

It is advantageous to transform $d_3$ and $f_3$ into operators acting on the
 solution space $Q_{\S_2}$ of the oscillator.  We have
\begin{eqnarray}
\D_3 & = & e^{\eta L_z}d_3e^{-\eta L_z} =
 \lfrac{1}{2}(\M_3+\L_z),\label{e:e40}\\*[1.5mm]
\F_3 & = & e^{\eta L_z}f_3e^{-\eta L_z} = \lfrac{1}{2}(\M_3-\L_z).\label{e:e41}
\end{eqnarray}
According to (\ref{e:apxbfea}), the spectra of these
operators are
\begin{eqnarray}
 & S(d_3) = S(\D_3) = \{m+\lfrac{1}{2}:m\in\bZp\} & \label{e:e42}\\*[1.5mm]
 & S(f_3) = S(\F_3) = \{n+\lfrac{1}{2}:n\in\bZp\}. & \label{e:e43}
\end{eqnarray}
Let $\Theta_{n,m}\in Q_{\S_2}$ be the coordinate representation
of the vector $\ketw{n}{m}$ in the representation space
$\uparrow_{-\frac{1}{2}}\otimes \uparrow_{-\frac{1}{2}}$.  Then, we have
\begin{eqnarray}
 & \F_3\Theta_{n,m} = (n+\lfrac{1}{2})\Theta_{n,m}, & \nonumber\\*[1.5mm]
 & \D_3\Theta_{n,m} = (m+\lfrac{1}{2})\Theta_{n,m}. & \label{e:e45}
\end{eqnarray}

We have already seen that the polar coordinate system $(\rho,\theta,\mu)$
is a natural coordinate system for this problem.  Therefore, we
express $\D_3$ and  $\F_3$ in these coordinates:
\begin{eqnarray}
\F_3 & = &
 \lfrac{i}{2}[\pt_3\partial_{\mu}-\lfrac{i}{4}\ddv_3\pt_3\rho^2+\lfrac{1}{2}
\dv_3+ih_0+\partial_{\theta}],\nonumber\\*[1.5mm]
\D_3 & = &
 \lfrac{i}{2}[\pt_3\partial_{\mu}-\lfrac{i}{4}\ddv_3\pt_3\rho^2+\lfrac{1}{2}
\dv_3+ih_0-\partial_{\theta}],\label{e:e47}
\end{eqnarray}
where $ \L_z = -i\partial_{\theta}$.
Substituting \ref{e:e47}) into (\ref{e:e45})
and solving the resulting first-order partial differential
equations, we obtain the result
\begin{equation}
\Theta_{n,m} = e^{i\R}\Upsilon_{n,m}(\rho)e^{i\theta(m-n)}T_{n,m}(\mu),
\label{e:e48}
\end{equation}
where $\R$ is given by (\ref{e:e22zb}) and
\begin{equation}
T_{n,m} = \varphi_3^{-\frac{1}{2}}\bigl (\frac{\bx}{\xi}\bigr
 )^{\frac{1}{2}(n+m+1)}
e^{-iK_0}.\label{e:e49}
\end{equation}
The radial functions $\Upsilon_{n,m}(\rho)$ are unknown at this point.
If we compare the functions (\ref{e:e22za}) with (\ref{e:e48}) and
(\ref{e:e11})
with (\ref{e:e49}), then we see that they are equivalent solutions if we
make the identification
\begin{equation}
u=n+m,~~l=m-n.\label{e:e49za}
\end{equation}

Note that these solutions are normed.  The norm is
\begin{eqnarray}
\int_{-\infty}^{+\infty}\int_{-\infty}^{+\infty}dxdy\bar{\Psi}_{n,m}
(x,y)\Psi_{n,m} (x,y) & = & \int_{-\infty}^{+\infty}\int_{-\infty}^{+\infty}
d\zeta_1d\zeta_2 \bar{\psi}_{0,0}(\zeta_1,\zeta_2)
\psi_{0,0}(\zeta_1,\zeta_2),\nonumber\\*[1.5mm]
   & = & \int_{0}^{\infty}\int_{0}^{2\pi}\rho d\rho d\theta
\bar{\Upsilon}_{n,m}(\rho)\Upsilon_{n,m}(\rho).\label{e:e49zc}
\end{eqnarray}

\subsection{Solution of the Radial Equation}

We can use the algebraic structure of $\G$ in the basis (\ref{e:e35za}) and the
structure of the irreducible
representation $\uparrow_{-\frac{1}{2}}\otimes \uparrow_{-\frac{1}{2}}$ to
obtain an explicit form for the radial functions.

Since the representation
$\uparrow_{-\frac{1}{2}}\otimes \uparrow_{-\frac{1}{2}}$ is bounded below
with respect to the spectra of both $f_3$ and $d_3$, there exists an
extremal state
\begin{eqnarray}
\Psi_{0,0}(\rho,\theta,\mu) & = & e^{-\eta L_z}\Theta_{0,0}(\rho,\theta,\mu)
\nonumber\\*[1.5mm]
           & = & e^{-\eta L_z}e^{i\R}\psi_{0,0}(\rho,\theta)T_{0,0}(\mu)
\nonumber\\*[1.5mm]
           & = & e^{-\eta L_z}e^{i\R}\Upsilon_{0,0}(\rho)T_{0,0}(\mu).
\label{e:e59zc}
\end{eqnarray}
Since
\begin{equation}
a_-\Psi_{0,0} = c_-\Psi_{0,0} = 0 \Rightarrow a\Upsilon_{0,0} = c\Upsilon_{0,0}
= 0,
\label{e:e59zd}
\end{equation}
we obtain
\begin{equation}
\Upsilon_{0,0} = \frac{1}{\sqrt{\pi}}e^{-\rho^2/2},\label{e:e64}
\end{equation}
after normalization.  Therefore, the extremal eigenstates are
\begin{eqnarray}
\Theta_{0,0} & = &
 \frac{1}{\sqrt{\pi}}e^{i\R}e^{-\rho^2/2},\nonumber\\*[2.0mm]
\Psi_{0,0} & = & \frac{1}{\sqrt{\pi}}e^{i\R}e^{-\eta
L_z}e^{-\rho^2/2}.\label{e:e66}
\end{eqnarray}
This means $\Theta_{0,0}=\Psi_{0,0}$.

According to Eqs. (\ref{e:apxbfea}), we have
\begin{equation}
\Psi_{n,m} =  \frac{1}{\sqrt{n!}\sqrt{m!}}a_+^nc_+^m\Psi_{0,0}.\label{e:e71}
\end{equation}
After some calculation \cite{drt4}, the complete wave function can be
written as
\begin{equation}
\Psi_{n,m} = e^{i\R}e^{-\eta L_z}\Upsilon_{n,m}(\rho)e^{i\theta(m-n)}
T_{n,m}(\mu),
\label{e:e79}
\end{equation}
in which the radial wave function $\Upsilon_{n,m}(\rho)$ is
\begin{equation}
\Upsilon_{n,m}(\rho) = \frac{(-)^k}{\sqrt{\pi}}\frac{k!}{\sqrt{n!}\sqrt{m!}}
L_k^{(|m-n|)}(\rho^2)\rho^{|m-n|}e^{-\rho^2/2},\label{e:e80zb}
\end{equation}
where
\begin{equation}
k=\lfrac{1}{2}[n+m-|m-n|]\label{e:e80}
\end{equation}
and the $\mu$-dependent function has the form
\begin{equation}
T_{n,m}(\mu) = \varphi_3^{-1/2}\bigl (\frac{\bx}{\xi}\bigr )^{(n+m+1)/2}
e^{-iK_0}. \label{e:e80za}
\end{equation}
The last equation may be compared to $T_u(\mu)$ in (\ref{e:e11}).
Note that for each $n$ there is an infinite degeneracy in $m$.

Recall that $m_3$ and $\L_z$ are diagonal in the basis $\{\Psi_{n,m}\}$,
since they are linear combinations of $d_3$ and $f_3$.
{}From (\ref{e:e36}), we see that
\begin{equation}
m_3 = d_3+f_3,~~~\L_z = d_3-f_3.\label{e:e81}
\end{equation}
Therefore, the eigenvalues of $m_3$ and $\L_z$ are
\begin{equation}
m_3\Psi_{n,m} = (m+n+1)\Psi_{n,m}\label{e:e82}
\end{equation}
and
\begin{equation}
\L_z\Psi_{n,m} = (m-n)\Psi_{n,m},\label{e:e83}
\end{equation}
where $n,m\in \bZp$.
This implies that  the spectrum of $m_3$ is
\begin{equation}
S(m_3) =\{m+n+1:n,m\in\bZp\},\label{e:e84}
\end{equation}
where each eigenvalue is $(n+m+1)$-fold degenerate.  The spectrum of $\L_z$ is
\begin{equation}
S(\L_z) = \{m-n:n,m\in\bZp\} = \{0,\pm 1,\pm 2, \cdots\},\label{e:e85}
\end{equation}
which is as expected since the operator $\L_z$ generates the group $O(2)$.
Each state can also be characterized by the eigenvalues of $m_3$ and $\L_z$.
Since each state with eigenvalue $n+m+1$ is $(n+m+1)$-fold degenerate, the
degenerate states are classified or labeled by the eigenvalues of $\L_z$.  They
are
\begin{equation}
\pm (m+n),~\pm (m+n-2),~\pm (m+n-4),~\cdots,~\pm 1~{\rm or}~0,\label{e:e86}
\end{equation}
according to whether $n+m+1$ is even or odd, respectively.

\newpage

%***********************************************************************
%\include{m6}
\section{Solutions to the Pauli Equation}

At this stage,
we can return to the original problem discussed in Sec. 1.
The solutions of the Schr\"odinger equation found in Sec. 5
and the symmetry structure developed in parallel
make possible the solution of the Pauli equation and
the identification of the supersymmetry.
In this section,
we present the solutions of the Pauli equation in a useful form.

The two-dimensional Pauli equation is reduced in Sec. 1 to
Eq. (\ref{e:a14}),
\begin{equation}
S_{2\pm}\Psi_\pm = 0~~,\label{e:f100}
\end{equation}
where $S_{2\pm}$ are given in
Eq. (\ref{e:a15}).
The symmetry analysis above for each of $S_{2+}$ and $S_{2-}$ yields
the same time-dependent coefficients
$\xi$, $\varphi_1$, $\varphi_2$, and $\varphi_3$.
Multiplying $S_{2\pm}$  by
$-\frac{1}{2}\varphi_3$ and using
$[a_-,a_+]=I$ in Eq. (\ref{e:d10}) and its $h_{0+} \rightarrow h_{0-}$ partner
gives
\begin{eqnarray}
-\frac{1}{2}\varphi_3S_{2\pm} & = & 2a_\pm a_\mp
+ \L_z - m_{3\pm} \pm 1, \nonumber\\*[1.5mm]
& = & 2a_\pm a_\mp - 2f_{3\pm} \pm 1 ,
\label{e:f13}
\end{eqnarray}
where $m_{3\pm}$ and $f_{3\pm}$
are given by (\ref{e:c26}) and (\ref{e:e36}) and their partners.
Equation (\ref{e:a14}) can therefore be rewritten as
\begin{equation}
(2a_ \pm a_\mp - 2f_{3\pm} \pm 1 ) \Psi_\pm = 0. \label{e:f16}
\end{equation}

{}From Sec. 5,
the solutions to this equation can be taken as
\begin{eqnarray}
\Psi_{\pm} & = & \Psi_{n_\pm,m_\pm}(x,y,\tau)\nonumber\\*[1.5mm]
              & = & e^{i\R}e^{-\eta L_z}\psi_{n_\pm,m_\pm}(\zeta_1,\zeta_2)
T_{\kappa_\pm;n_\pm,m_\pm}(\mu), \label{e:f17}
\end{eqnarray}
where $\Psi_{n_+,m_+}(x,y,\tau)$ is given by (\ref{e:e79}) and
$\Psi_{n_-,m_-}(x,y,\tau)$ is its partner.
The function $\psi_{n,m}$ is a
function of either of the variables of separation $(\zeta_1,\zeta_2)$ or
$(\rho,\theta)$.
The result is a double eigenvalue problem:
\begin{equation}
a_\pm a_\mp \Psi_{n_\pm,m_\pm} =
\left( \begin{array}{c} \scriptstyle n_+ \\ \scriptstyle n_-+1 \end{array}
\right)
\Psi_{n_\pm,m_\pm}.
\end{equation}
Using (\ref{e:f17}) for the solution $\Psi_{n_+,m_+}$ and
a similar expression for its partner,
and noting (\ref{e:e31}) and (\ref{e:e32}), we obtain the result
\begin{equation}
A_\pm A_\mp \psi_{n_\pm,m_\pm}T_{n_\pm,m_\pm} =
\left( \begin{array}{c} \scriptstyle n_+ \\ \scriptstyle{n_-+1} \end{array}
\right)
\psi_{n_\pm,m_\pm}T_{\kappa_\pm;n_\pm,m_\pm}.\label{e:f20}
\end{equation}

Finally, since the operators $A_{\pm}$ contain no time derivatives and
because of Eqs. (\ref{e:e59}),
the two-component Pauli equation can be written as
\begin{equation}
\left( \begin{array}{cc}
\scriptstyle \adg a  & \scriptstyle 0 \\
\scriptstyle 0 & \scriptstyle a\adg \end{array} \right) \left( \begin{array}{c}
\scriptstyle \psi_{n_+,m_+} \\
\scriptstyle \psi_{n_-,m_-} \end{array} \right) = \left( \begin{array}{cc}
\scriptstyle n_+ & \scriptstyle 0 \\
\scriptstyle 0 & \scriptstyle{n_-+1} \end{array}
 \right)\left( \begin{array}{c}
\scriptstyle \psi_{n_+,m_+} \\
\scriptstyle \psi_{n_-,m_-} \end{array} \right),
\label{e:f21}
\end{equation}
where the operators $\adg a$ and $a\adg$ are
given in the coordinate representation by
\begin{eqnarray}
\adg a & = &
\lfrac{1}{4}[-(\partial_{\zeta_1\zeta_1}+\partial_{\zeta_2\zeta_2})
 -2i(\zeta_2\dzo-\zeta_1\dzw) +(\zeta_1^2+\zeta_2^2) -2],\nonumber\\*[1.5mm]
a\adg & = & \lfrac{1}{4}[-(\partial_{\zeta_1\zeta_1}+\partial_{\zeta_2\zeta_2})
 -2i(\zeta_2\dzo-\zeta_1\dzw) +(\zeta_1^2+\zeta_2^2) +2].\label{e:f23}
\end{eqnarray}
Equation (\ref{e:f21}) is the generalization of the factorization obtained in
Ref. \cite{fknt1}.

\newpage

%*********************************************************************
%\include{m7}
\section{Supersymmetry}

In this section,
we identify a supersymmetry associated with a
nonrelativistic charged spin-$\lfrac{1}{2}$ particle
moving in a time-varying uniform magnetic induction
${\bf B}(\tau) = B(\tau) {\hat{\bf z}}$.
This supersymmetry generalizes that discussed in
Refs. \cite{jackiw,hkn,fknt1}.

The relevant time-dependent Pauli equation for this system
is Eq. (\ref{e:a1}),
with $\phi({\bf r},t)=0$ and the gauge choice
in Eq. (\ref{e:a4}).
For more generality,
we do not take $p_z$ as zero but
instead separate variables with respect to $z$.
The momentum in the $z$-direction is then represented
in the expressions below by its eigenvalue $\kappa_\pm$.

As described in the previous sections,
the solution space $Q_{S_{2+}}$ spanned by the functions
$\{\Psi_{n_+,m_+}: n_+,m_+\in \bZp\}$ forms a basis for a representation space
for
the irreducible representation
$\uparrow_{-\lfrac{1}{2}}\otimes\uparrow_{-\lfrac{1}{2}}$
of the Lie algebra of operators $\{ d_{3+}, f_{3+}, a_{\pm}, c_{\pm},I\}$.
A partner space $Q_{S_{2-}}$ spanned by the partner functions
$\{\Psi_{n_-,m_-}: n_-,m_-\in \bZp\}$
also exists and forms a basis for the same irreducible representation
of the isomorphic Lie algebra of operators
$\{ d_{3-}, f_{3-}, a_{\pm}, c_{\pm},I\}$.
The spectra are
\beq
  S(f_{3\pm})= \{ n_\pm +\lfrac{1}{2}: n_\pm\in \bZp\},
{}~~S(d_{3\pm}) = \{ m_\pm +\lfrac{1}{2}: m_\pm\in \bZp\}. \label{e:g4}
\eeq
{}From Eq. (\ref{e:f21})
we see that the spectrum of the operator $\adg a$ is
\begin{equation}
S(\adg a) = \{n_+:n_+ \in \bZp\} = \{0,1,2,\cdots\}, \label{e:g6}
\end{equation}
while the spectrum of $a\adg$ is
\begin{equation}
S(a\adg) = \{n_-+1:n_-\in \bZp\} = \{1,2,3\cdots\}. \label{e:g7}
\end{equation}
In effect, the two operators $\adg a$ and $a\adg$ have
identical spectra, except that the latter is missing the ground state.
This suggests the existence of a supersymmetry.

Following Ref. \cite{fknt1},
we introduce a unified notation that permits the simultaneous handling
of the two spaces.
We define a parameter $\nu$ that takes the value 0 for the
`bosonic' space $Q_{S_{2+}}$ and 1 for the `fermionic' space $Q_{S_{2-}}$.
It distinguishes the upper and lower components of the two-component
Pauli equation.
States in the two spaces can then be denoted by
$\kettsc{n}{m}{\nu}$,
where $n = 0,1,2,\ldots$ and $m +\half \in \bZp$.
For the bosonic space $n = n_+$ and $m = m_+$, while for the
fermionic space $n = n_-$ and $m = m_-$.
In two-component notation,
we have
\begin{equation}
\kettsc{n}{m}{\nu} = \psi_{n,m}(\zeta_1,\zeta_2)\left(\begin{array}{c}
\delta_{0\nu} \\ \delta_{1\nu}\end{array}\right).\label{e:g11}
\end{equation}
The action of the operators $a$, $c$ and their conjugates becomes
\begin{eqnarray}
 & a\kettsc{n}{m}{\nu} = \sqrt{n}\kettsc{n-1}{m}{\nu},~~\adg\kettsc{n}{m}{\nu}
=
\sqrt{n+1}\kettsc{n+1}{m}{\nu}, & \nonumber\\*[1.5mm]
 & c\kettsc{n}{m}{\nu} = \sqrt{m}\kettsc{n}{m-1}{\nu},~~\cdg\kettsc{n}{m}{\nu}
=
\sqrt{m+1}\kettsc{n}{m+1}{\nu}, & \label{e:g15}
\end{eqnarray}
These expressions are the natural time-varying extensions of the
results obtained in Ref. \cite{fknt1}.

We can also define raising and lowering operators $b$ and $\bdg$
for the index $\nu$.
By definition,
these operators satisfy the anticommutation relations
\begin{equation}
\{b,\bdg\} = I,~~\{b,b\} = \{\bdg,\bdg\} = 0.
\label{e:g2}
\end{equation}
Their two-component form is
\begin{equation}
 b=\left( \begin{array}{cc} 0 & 1\\ 0 & 0\end{array}\right),~~\bdg =
\left( \begin{array}{cc} 0 & 0\\ 1 & 0 \end{array}\right). \label{e:g1}
\end{equation}
Their action on the ket $\kettsc{n}{m}{\nu}$ is given by
\begin{equation}
b\kettsc{n}{m}{\nu} = \delta_{1\nu}\kettsc{n}{m}{0},~~\bdg\kettsc{n}{m}{\nu} =
\delta_{0\nu}\kettsc{n}{m}{1}. \label{e:g16}
\end{equation}

We can now introduce the time-dependent operator $\hat H$,
defined by
\beq
\hat H =  \adg a + \bdg b.
\label{e:g100}
\end{equation}
Note that, although $\hat H$ is an integral of the motion,
there is implicit time dependence in $a$ and $\adg$.
When normalized as $H = eB\hat H/M$,
this operator is a time-dependent extension  of
the usual hamiltonian for the time-independent Landau problem.
It directly generalizes the operator denoted by $\hat H$
in Ref. \cite{fknt1}.
Using Eq. (\ref{e:g100}) and the new ket notation,
the factorized Schr\"odinger equation (\ref{e:f21})
becomes
\beq
\hat H\kettsc{n}{m}{\nu}  = (n+\nu) \kettsc{n}{m}{\nu} .
\label{e:g101}
\end{equation}
This expression shows that the states
$\kettsc{n_+}{m_+}{0}$ and $\kettsc{n_-=n_+-1}{m_-}{1}$ are degenerate,
except for the unique ground state $\kettsc{0}{0}{0}$,
which has zero eigenvalue for $\hat H$
(as required for unbroken supersymmetry).

The framework for the supersymmetry is now almost complete.
It remains merely to introduce appropriate supersymmetry generators
mapping degenerate states into one another.
These operators are defined by
\begin{equation}
Q = a\bdg,~~~\Qdg =\adg b.\label{e:g8}
\end{equation}
They satisfy the graded commutation
relations of the supersymmetric quantum-mechanical algebra sqm(2):
\begin{eqnarray}
 & \{Q,Q\} = \{\Qdg,\Qdg\} = 0,~~~\{Q,\Qdg\} =  \hat{H}, & \nonumber\\*[1.5mm]
 & [\hat{H},Q] = [\hat{H},\Qdg] = 0. & \label{e:g10}
\end{eqnarray}
Explicitly, the action of the supersymmetry generators is
\begin{eqnarray}
 & Q\kettsc{n}{m}{\nu} =\sqrt n  \delta_{0\nu}\kettsc{n-1}{m}{1},
 & \nonumber\\*[1.5mm]
 & \Qdg\kettsc{n}{m}{\nu} = \sqrt{n+1}\delta_{1\nu}\kettsc{n+1}{m}{0}.
 & \label{e:g18}
\end{eqnarray}
The operator $Q$ maps bosonic states into fermionic ones,
while its conjugate does the reverse.

\newpage

%**********************************************************************
%\include{m8}
\section{Supercoherent States}

At this stage,
we are in a position to construct the supercoherent states of the time-varying
Landau system.
To do so requires a supersymmetric generalization of the standard
displacement-operator method.
A natural approach to this was introduced in
Ref. \cite{fknt1},
using the supermanifold formalism developed in
Ref. \cite{rog} and the techniques for BCH relations
of Ref. \cite{knt1}.
For a description of the general procedure and examples
of its application,
see the above references and Refs. \cite{kr,tkn1}.

In the present case,
the relevant superalgebra $\G_s$ is the one
obtained by extending the Lie algebra $\G$ as follows:
\begin{equation}
\G_s = \{\adg a,\cdg c, a, \adg, c, \cdg, 1; \bdg b, b, \bdg\}.\label{e:ga1}
\end{equation}
A fixed state is required by the construction.
We choose it as the ground state
$\kettsc{0}{0}{0}$.
The subalgebra consisting of the operators
$\{\adg a,\cdg c, 1; \bdg b\}$ leaves this state fixed,
i.e., the ground state is an eigenvector of these operators.
According to the procedure of Ref. \cite{fknt1},
the supercoherent states are
to be defined via the action of the operators in the quotient algebra
$\{a, \adg, c, \cdg; b, \bdg\}$ on the fixed state.

It is convenient to define a super-Hermitian basis for the quotient algebra:
\begin{eqnarray}
 & X_1 = a+\adg,~~~X_2 = i(a-\adg),~~~X_3 = c+\cdg,~~~X_4 = i(c-\cdg), &
\nonumber\\*[1.5mm]
 & X_5 = i(b+\bdg),~~~X_6 = \bdg-b. & \label{e:ga3}
\end{eqnarray}
Constructing a unitary representation of the supergroup for an element in the
quotient algebra \cite{fknt1,tkn1}, we obtain
\begin{eqnarray}
T(g) & = & \exp{(iA_1X_1+iA_2X_2+iC_3X_3+iC_4X_4+i\theta_1X_5+i\theta_2X_6)},
\nonumber\\*[1.5mm]
     & = & \exp{(-\bar{A}a + A\adg -\bar{C}c + C\cdg +\theta \bdg +
\bar{\theta}b)},\label{e:ga4}
\end{eqnarray}
where $A= A_2+iA_1$, $C = C_2+iC_1$, and $\theta = -\theta_1+i\theta_2$.  This
means that $A,C\in \,^0B_L$ and $\theta\in \,^1B_L$
are complex Grassmann-valued variables.

Using a suitable Baker-Campbell-Hausdorff relation for the supergroup
element and Lemma 1 of Ref. \cite{knt1},
we can construct the analogues of the
supercoherent states in Ref. \cite{fknt1}.
The states are parametrized by
three Grassmann-valued parameters $A$, $C$, and $\theta$,
and are given as
\begin{eqnarray}
\vert Z\rangle & = & \exp{(\lfrac{1}{2}\theta\bar{\theta})}
\exp{(-\lfrac{1}{2}|A|^2)}\exp{(-\lfrac{1}{2}|C|^2)}\nonumber\\*[1.5mm]
               &   &~~~~~~\sum_{n,m}{{A^nC^m}\over{\sqrt{n!}\sqrt{m!}}}
(\kettsc{n}{m}{0}+\theta\kettsc{n}{m}{1}).\label{e:ga5}
\end{eqnarray}
Recall that the operators $a$, $\adg$, $c$, and
$\cdg$ have an implicit time-dependence in them through their dependence upon
the variables $\zeta_1=x/\varphi_3^{1/2}({\tau})$ and
$\zeta_2=y/\varphi_3^{1/2}({\tau})$.
The supercoherent states defined above
are the natural time-dependent generalizations of the
supercoherent states derived in \cite{fknt1}.

In Ref. \cite{m12},
several theorems about time-dependent integrals of the motion are given.
The first states that any function of integrals of the motion is
itself an integral of the motion.
The second states that eigenvalues of time-dependent integrals of
the motion do not depend on time.
The third states that applying an integral of the motion on a
solution to a wave equation
(either a Schr\"odinger or a Pauli time-dependent equation) yields a
function that itself is a solution to the same wave equation.
Since $a$, $b$,
$c$, and their conjugates are integrals of the motion,
so too is the operator $T(g)$.
Furthermore, since $\vert 0,0,0\rangle$ is a solution to the Pauli equation,
$\vert Z\rangle$ is also a solution to the same equation.
All the formulae in Ref. \cite{fknt1}
may be rewritten identically for the properties of the present supercoherent
states, with obvious replacements.
Note that some aspects of integrals of motion in simple problems
with supersymmetric quantum mechanics have been discussed in
Ref. \cite{m14}.

If one introduces a time-dependent electric field
in addition to the present magnetic field,
the basic algebraic structure derived here can again be applied.
Provided this leads to a supersymmetric formulation,
this provides a means of obtaining the explicit form of supercoherent states
for this more general case.
It is plausible that supercoherent
states for a relativistic electron moving in an electromagnetic field
of arbitrary configuration could be found using
these methods combined with coherent states in the proper-time formalism,
as discussed in Ref. \cite{m13}.

\newpage

%***************************************************************
%\include{m9}
\section{Appendix: Constant Magnetic Induction}

In this appendix,
we demonstrate explicitly the connection between our results
and the special case of constant induction treated in Ref. \cite{fknt1}.

When the magnetic induction $B$ is constant and uniform,
$w = eB$,
the differential equations (\ref{e:b19}) and (\ref{e:b20}) take the form
\begin{equation}
\ddot{\epsilon}^{\alpha}+\lfrac{1}{4}w^2\epsilon^{\alpha} = 0,~~~~\alpha = 1,2,
\label{e:Ap-c}
\end{equation}
and have real solutions
\begin{equation}
\chi_1 = \sqrt{\frac{2}{w}}\cos(\frac{w\tau}{2}),
{}~~~\chi_2 = \sqrt{\frac{2}{w}}\sin(\frac{w\tau}{2}) \label{e:Ap-d}
\end{equation}
satisfying the wronskian condition $W(\chi_1,\chi_2) = 1$.
The complex solutions (\ref{e:c2}) become
\begin{equation}
\xi = \frac{1}{\sqrt{2}}(\chi_1+i\chi_2) = \frac{1}{\sqrt{w}}e^{iw\tau /2},
{}~~~\bx = \frac{1}{\sqrt{w}}e^{-iw\tau /2}, \label{e:Ap-e}
\end{equation}
with wronskian $W(\xi,\bx) = -i$.

The complex solutions to the equation
(\ref{e:b18}) are given by (\ref{e:c5}):
\begin{equation}
\varphi_1 = \frac{1}{w}e^{iw\tau}, ~~~\varphi_2 = \frac{1}{w}e^{-iw\tau},
{}~~~\varphi_3 = \frac{2}{w},~~~\eta = \frac{1}{2}w\tau. \label{e:Ap-f}
\end{equation}

Using the equations derived in Sec. 5.2, we
can write down the generators of the Schr\"odinger algebra
\begin{eqnarray}
m_- & = & \frac{1}{w}e^{iw\tau}[-\partial_{\tau} - \frac{w}{2}L_z
-\frac{iw}{2}(x\partial_x+y\partial_y) -\frac{i}{4}w^2(x^2+y^2) - \frac{iw}{2}
-ih_0], \nonumber \\*[1.5mm]
m_+ & = & \frac{1}{w}e^{-iw\tau}[\partial_{\tau} + \frac{w}{2}L_z
-\frac{iw}{2}(x\partial_x+y\partial_y) +\frac{i}{4}w^2(x^2+y^2) - \frac{iw}{2}
+ih_0], \nonumber \\*[1.5mm]
m_3 & = & \frac{2i}{w}[\partial_{\tau} + \frac{1}{w}L_z + ih_0],
\nonumber \\*[1.75mm]
\L_z & = & iL_z = i(y\partial_x-x\partial_y), \nonumber\\*[1.75mm]
a_- & = & \frac{e^{iw\tau}}{\sqrt{2w}}[\partial_x+i\partial_y
+ \frac{w}{2}(x+iy)], \nonumber \\*[1.5mm]
a_+ & = & \frac{e^{-iw\tau}}{\sqrt{2w}}[-(\partial_x-i\partial_y)
+\frac{w}{2}(x-iy)], \nonumber \\*[1.5mm]
c_- & = & \frac{1}{\sqrt{2w}}[\partial_x-i\partial_y +\frac{w}{2}(x-iy)],
\nonumber \\*[1.5mm]
c_+ & = & \frac{1}{\sqrt{2w}}[-(\partial_x+i\partial_y)+\frac{w}{2}(x+iy)],
\nonumber \\*[1.5mm]
I & = & 1. \label{e:Ap-m}
\end{eqnarray}
The generators of the subalgebra $\G$ are
$\G =\{d_3,f_3,a_{\pm},c_{\pm},I\}$. We have
\begin{eqnarray}
f_3 & = & \frac{1}{w}(i\partial_{\tau} -h_0), \nonumber\\*[1.5mm]
d_3 & = & \frac{1}{w}(i\partial_{\tau} + w\L_z - h_0),\label{e:Ap-o}
\end{eqnarray}
along with the operators (\ref{e:Ap-m}).

{}From the results of Section 5, the solutions (\ref{e:e79}) are
\begin{equation}
\Psi_{n,m} = e^{-\eta L_z}\Upsilon_{n,m}(\rho)e^{i\theta(m-n)}T_{n,m}(\mu),
\label{e:Ap-r}
\end{equation}
where $\Upsilon_{n,m}(\rho)$ is given by (\ref{e:e80zb}) and
\begin{equation}
T_{n,m} = (\lfrac{w}{2})^{\lfrac{1}{2}}e^{-i(n+m+1)w\mu/2}e^{-ih_0\mu},
\label{e:Ap-s}
\end{equation}
where $n,\,m\in \bZp$.
The polar coordinate system is related to the cartesian system through the
transformation
\begin{equation}
\rho^2 = \zeta_1^2+\zeta_2^2,~~~\theta = \tan^{-1}{\zeta_2/\zeta_2},
\label{e:Ap-t}
\end{equation}
where
\begin{equation}
\zeta_1 = x(\frac{w}{2})^{\lfrac{1}{2}},~~\zeta_1 =
x(\frac{w}{2})^{\lfrac{1}{2}},
{}~~\mu=\tau,\label{e:Ap-u}
\end{equation}
following Eqs. (\ref{e:e8}).

By direct calculation, we see that the action of $f_3$ and $d_3$ on
$\Psi_{n,m}$ is
\begin{eqnarray}
f_3\Psi_{n,m} & = & (n+\lfrac{1}{2})\Psi_{n,m},\nonumber\\*[1.5mm]
d_3\Psi_{n,m} & = & (m+\lfrac{1}{2})\Psi_{n,m}.\label{e:Ap-w}
\end{eqnarray}
Also, the ladder operators $a_{\pm}$ and $c_{\pm}$ raise and lower the quantum
numbers $n$ and $m$, respectively.  Thus
\begin{eqnarray}
 & a_-\Psi_{n,m} = \sqrt{n}\Psi_{n-1,m},~~a_+\Psi_{n,m} =
\sqrt{n+1}\Psi_{n+1,m}, &
\nonumber\\*[1.5mm]
 & c_-\Psi_{n,m} = \sqrt{m}\Psi_{n,m-1},~~c_+\Psi_{n,m} =
\sqrt{m+1}\Psi_{n,m+1}. &
\label{e:Ap-y}
\end{eqnarray}

In the special case of constant field, the $\tau$ or $\mu$ variable may be
separated and we can work with the cartesian-like coordinates
($\zeta_1,\,\zeta_2$) or with the polar coordinates ($\rho,\,\theta$), as
defined in (\ref{e:Ap-u}) and (\ref{e:Ap-t}), respectively.  In the former
case, use the wavefunction $\psi_{n,m}(\zeta_1,\zeta_2)$, and in the latter
we use $\psi_{n,m}(\rho,\theta)=\Upsilon_{n,m}(\rho)e^{i\theta(m-n)}$.
On these solution spaces, the operators $f_3$ and $d_3$ are replaced by
$N_a=\adg a$ and $N_c=\cdg c$, respectively.  The ladder operators $a$
and $c$ are defined by
\begin{eqnarray}
 a & = & \lfrac{1}{2}[(\dzp) + (\zpz)],\nonumber\\*[1.5mm]
   & = & \lfrac{1}{\sqrt{2w}}[(\partial_x+i\partial_y) +\lfrac{w}{2}(x+iy)],
\nonumber\\*[1.5mm]
%\adg & = & \lfrac{1}{2}[-(\dzm) + (\zmz)],\nonumber\\*[1.5mm]
%     & = & \lfrac{1}{\sqrt{2w}}[-(\partial_x-i\partial_y)
%%+\lfrac{w}{2}(x-iy)],
%\label{e:Ap-a1}\\*[1.5mm]
 c & = & \lfrac{1}{2}[(\dzm) + (\zmz)], \nonumber\\*[1.5mm]
   & = & \lfrac{1}{\sqrt{2w}}[(\partial_x-i\partial_y) +\lfrac{w}{2}(x-iy)],
\label{e:Ap-b1}%\\*[1.5mm]
%\cdg & = & \lfrac{1}{2}[-(\dzp) + (\zpz)], \nonumber\\*[1.5mm]
%     & = & \lfrac{1}{\sqrt{2w}}[-(\partial_x+i\partial_y)
%%+\lfrac{w}{2}(x+iy)],
%\label{e:Ap-c1}
\end{eqnarray}
where we have made use of (\ref{e:Ap-u}).  In polar coordinates (\ref{e:Ap-t}),
the operators and their conjugates are given in Eqs. (\ref{e:e57}).

The operators in (\ref{e:Ap-b1}) are exactly the $a$ and $c$
operators obtained in Ref. \cite{fknt1}.  Their time-dependent
extensions are $a_-$ and $c_-$, respectively.

The number operators
obey the nonzero commutation relations
\begin{eqnarray}
 & [N_a,a] = -a,~~[N_a,\adg] = +\adg,~~[a,\adg] = I, & \nonumber\\*[1.5mm]
 & [N_c,c] = -c,~~[N_c,\cdg] = +\cdg,~~[c,\cdg] = I. & \label{e:Ap-e1}
\end{eqnarray}
The action of these operators on the manifold of states $\{\psi_{n,m}\}$ is
\begin{eqnarray}
 & N_a\psi_{n,m} = n\psi_{n,m},~~a\psi_{n,m}=\sqrt{n}\psi_{n-1,m},~~
\adg\psi_{n,m} = \sqrt{n+1}\psi_{n+1,m}, & \nonumber\\*[1.5mm]
 & N_c\psi_{n,m} = m\psi_{n,m},~~c\psi_{n,m}=\sqrt{m}\psi_{n,m-1},~~
\cdg\psi_{n,m} = \sqrt{m+1}\psi_{n,m+1}. & \label{e:Ap-g1}
\end{eqnarray}
Substituting (\ref{e:Ap-b1}) for $a$ in $N_a$, we
obtain
\begin{eqnarray}
N_a & = & \lfrac{1}{4}[-(\partial_{\zeta_1\zeta_1}+(\partial_{\zeta_2\zeta_2}
+2\L_z) + (\zeta_1^2+\zeta_2^2)-2]\nonumber\\*[1.5mm]
    & = & \lfrac{1}{2w}[-(\delw +w\L_z) + \lfrac{w^2}{4}(x^2+y^2)-w],
\label{e:Ap-i1}
\end{eqnarray}
which is proportional to the Hamiltonian \cite{fknt1}.

This completes the explicit demonstration that our formalism
contains as a special case the description of the motion of a nonrelativistic
spin-$\lfrac{1}{2}$ charged particle in a constant and uniform magnetic
induction.
\newpage

%***********************************************************
%\include{m10}
\section{Acknowledgements}

This research was supported in part by the United States Department of
Energy under contracts DE-AC02-84ER40125 and DE-FG02-91ER40661 (V.A.K.),
W-7405-ENG-36 (M.M.N.), and by the Natural Sciences and Engineering
Research Council of Canada (D.R.T.).

\end{document}